\documentclass{aa}  

\usepackage{graphicx}
\usepackage{txfonts}
\usepackage{natbib}
\usepackage{hyperref}
\usepackage{arydshln}

\begin{document}

   \title{Insights into the chemical composition of the metal-poor Milky Way halo globular cluster \object{NGC~6426}\thanks{This  paper  includes  data  gathered  with  the  6.5-m  Magellan
          Telescopes located at Las Campanas Observatory, Chile.}
         }

   \author{M. Hanke\inst{1,2}
          \and
          A. Koch\inst{1,3}
          \and
          C. J. Hansen\inst{4}
          \and
          A. McWilliam\inst{5}
          }

   \institute{Landessternwarte, Zentrum für Astronomie der Universität Heidelberg, Königstuhl 12, 69117 Heidelberg, Germany
              \and
              Astronomisches  Rechen-Institut,  Zentrum  für  Astronomie  der  Universität  Heidelberg,  Mönchhofstr.
12-14,  69120  Heidelberg, Germany\\
              \email{mhanke@ari.uni-heidelberg.de}
              \and
              Department of Physics, Lancaster University, Bailrigg, Lancaster LA1 4YB, UK
              \and
              Dark Cosmology Centre, The Niels Bohr Institute, Juliane Maries Vej 30, 2100 Copenhagen, Denmark
              \and
              Carnegie Observatories, 813 Santa Barbara St., Pasadena CA 91101, USA
             }

   \date{Received 5 September 2016 /
         Accepted 7 December 2016}
 
\abstract{We present our detailed spectroscopic analysis of the chemical composition of four red giant stars in the halo globular cluster NGC~6426. We obtained high-resolution spectra using the Magellan2/MIKE spectrograph, from which we derived equivalent widths and subsequently computed abundances of 24 species of 22 chemical elements. For the purpose of measuring equivalent widths, we developed a new semi-automated tool, called EWCODE. We report a mean Fe content of [Fe/H$]=-2.34\pm0.05$ dex (stat.) in accordance with previous studies. At a mean $\alpha$-abundance of [(Mg,Si,Ca)/3 Fe$]=0.39\pm0.03$ dex, NGC~6426 falls on the trend drawn by the Milky Way halo and other globular clusters at comparably low metallicities. The distribution of the lighter $\alpha$-elements as well as the enhanced ratio [Zn/Fe$]=0.39$ dex could originate from hypernova enrichment of the pre-cluster medium. We find tentative evidence for a spread in the elements Mg, Si, and Zn, indicating an enrichment scenario, where ejecta of evolved massive stars of a slightly older population have polluted a newly born younger one. The heavy element abundances in this cluster fit well into the picture of metal-poor globular clusters, which in that respect appear to be remarkably homogeneous. The pattern of the neutron-capture elements heavier than Zn points toward an enrichment history governed by the r-process with little, if any, sign of s-process contributions. This finding is supported by the striking similarity of our program stars to the metal-poor field star \object{HD~108317}.}
\titlerunning{The chemical composition of the globular cluster NGC~6426}
\maketitle

\section{Introduction}
Amongst the oldest stellar systems known to exist in the Milky Way (MW) are metal-poor globular clusters (GCs). These accumulations of stars do not seem to have undergone substantial star formation for extended periods. Given the limited quality of the available data, for a long time color-magnitude diagrams (CMDs) of GCs appeared to be narrow and could be readily described by a single isochrone. These observations have justified the establishment of the long-lasting paradigm that considers CGs as prime examples of simple stellar populations (SSPs), that is, the results of very short bursts of star formation in their natal clouds. However, improved photometric precision indicates the presence of sub-populations in the cluster CMDs that are inconsistent with the SSP assumption, for a number of luminous GCs in a variety of bandpasses. Thus, early detections of chemical abundance variations \citep[e.g.,][]{Cohen78} could be more easily explained in a scenario involving several populations. Moreover, in recent years evidence has grown supporting the statement that GCs are generally composed of two or three chemically distinct populations. These subpopulations are separated by a few tens to hundreds of Myr in age and show vastly varying abundances of light elements such as C, N, O, Na, Mg, and Al \citep[see, e.g.,][and references therein]{Carretta09,Gratton12}. Theoretical considerations \citep[see, e.g.,][]{Dercole08,Dercole11} imply that GCs could have lost the majority of the initial stellar content of the first population, which consequently should have ended up in the Galactic halo. In fact, numerous studies found metal-poor GCs to be consistent with the abundance trends of the MW halo at equally low metal content \citep[e.g.,][]{Pritzl05,Koch09,Koch14,Villanova16}. We address this scenario by adding NGC~6426 to the short list of metal-poor clusters with available information on detailed chemical abundances. There are only two GCs in the Harris catalog \citep[][2010 edition]{Harris96} more metal poor than NGC~6426. At $12.9\pm1.0$ Gyr, the cluster is the oldest in the age compilation by \citet{Salaris02}. At a galactocentric distance of $R_\mathrm{gc}=14.4$ kpc and a galactic latitude of $16.23^{\circ}$ it is located in the transition region between inner and outer halo. Previous studies found consistent [Fe/H]\footnote{In this paper, we have employed the common abundance notation $\log{\epsilon}(\mathrm{X})=\log{(n(\mathrm{X})/n(\mathrm{H}))}+12$ for the element X. The bracket notation $\mathrm{[ A / B ]}$ relates the stellar abundances to the respective solar ratios for the elements A and B taken from \citet{Asplund09}.} values: $-2.20\pm0.17$ dex \citep{Zinn84}, $-2.33\pm0.15$ \citep{Hatzidimitriou99}, and $-2.39\pm0.04$ dex \citep{Dias15}. The latter value originates from the very first spectroscopic analysis of NGC~6426 at low resolution, which also stated [Mg/Fe$]=0.38\pm0.06$ dex. To date, there has been no study further addressing the detailed metal content of this cluster.

In this work, we have derived chemical abundances of 22 elements from high resolution spectroscopy of four giant stars attributed to NGC~6426. For the first time, we detected elements heavier than Fe in NGC~6426 and identify potential sites of nucleosynthesis that contributed to their enrichment. Furthermore, we relate NGC~6426's chemical composition to the framework of other GCs and the MW halo.  

This paper starts by introducing the sample and basic steps of data gathering and subsequent reduction (Sect. \ref{Sec:obs+red}). Next, we introduce the analysis methods applied, including a brief introduction to the newly developed semi-automated equivalent width code, EWCODE (Sect. \ref{Sec:abanalysis}). There too, we describe the methods applied in order to derive the basic stellar photospheric parameters entering our model atmospheres. In Sect. \ref{Sec:abres}, the results of this study are presented. Finally, Sect. \ref{Sec:summ} recapitulates all essential findings and the conclusions we draw from them.

\section{Observations and data reduction}\label{Sec:obs+red}
\begin{figure}
  \centering
  \resizebox{\hsize}{!}{\includegraphics{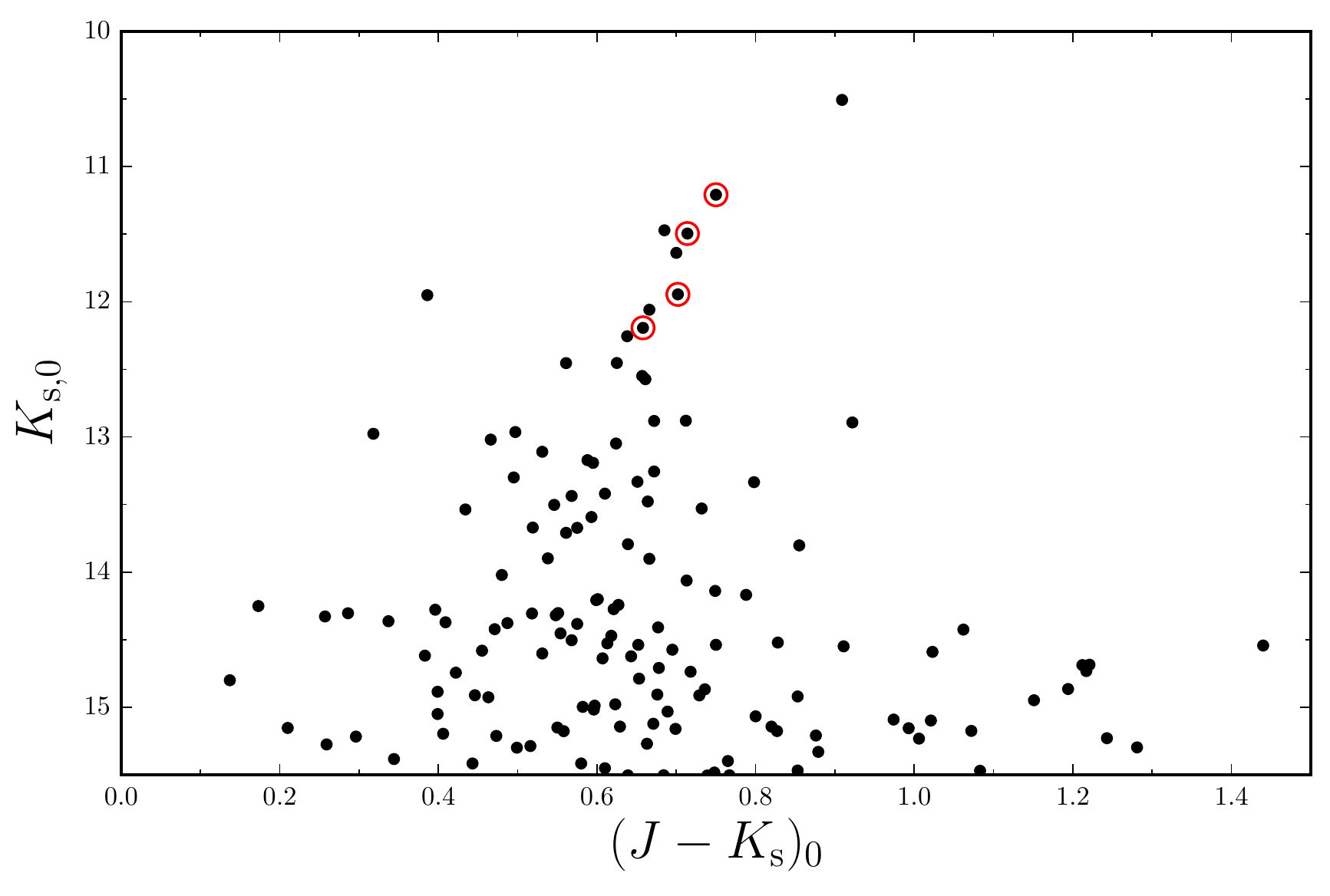}}
     \caption{Dereddened $(J-K_\mathrm{s})_0$ vs. 2MASS $K_{\mathrm{s},0}$ \citep{Schlafly11} color-magnitude diagram (CMD) of NGC~6426. Black dots correspond to stars within two half light radii and open red circles highlight the sample analyzed in this study.
              }
     \label{Fig:CMD}
  \end{figure}

  Our four NGC~6426 target stars were selected from what we identified as the tip of the red giant branch (RGB) in a 2MASS \citep{2MASS} $K_\mathrm{s}$ vs. $J-K_\mathrm{s}$ CMD covering the region of two half-light radii ($\sim0.92\arcmin$) around the cluster's center at $\alpha=$ 17:44:54.65, $\delta=$ +03:10:12.5 \citep[J2000,][]{Harris96}. The selection is indicated in Fig. \ref{Fig:CMD}. We decided against the usage of the recent Hubble Space Telescope (HST) photometry by \citet{Dotter11}, because the bright stars in question were labeled with bad fit quality flags and are thus likely to have saturated in the imaging process. As a consequence, only three of our targets have available HST $V$ and $I$ magnitudes. 
  
  \subsection{Data acquisition and reduction}  
  The spectra were gathered over three nights in August 2014 and August 2015 using the Magellan Inamori Kyocera Echelle (MIKE) spectrograph \citep{Bernstein03}, which is mounted at the Clay/Magellan2 Telescope at Las Campanas Observatory in Chile. We chose a slit width of 0.5$\arcsec$ together with an on-chip 2$\times$1 binning. Using this configuration and both the blue and red arms of the instrument, we ended up with full wavelength coverage in the range $3340 \mathrm{  \AA} \leq \lambda \leq 9410 \mathrm{  \AA}$. From the full widths at half maximum ($FWHM$) of the calibration lamp emission lines we derived a median spectral resolving power $R=\lambda/ FWHM \sim 45\,000,$ and $\sim 40\,000$ for the blue and the red arms respectively. The average seeing ranged from 0.6$\arcsec$ for the star 5659 up to 0.9$\arcsec$ for 2314. The basic observational settings are provided in Table \ref{Table:TARGET_LIST}. 
  
  The data were processed with the pipeline reduction package of \citet{Kelson03}, which takes into account flat field division, order tracing from quartz lamp flats, and wavelength calibrations using built-in Th-Ar lamp spectra, that were obtained immediately following each science exposure.
  
  \begin{table*}
  \caption{Observing information}
  \label{Table:TARGET_LIST}
  \centering
  \begin{tabular}{l c c c c c c c}
  \hline\hline
  Star-ID\tablefootmark{a} & $\alpha$ & $\delta$ & Date of observation & Exposure time & avg. seeing & S/N\tablefootmark{b} & $v_{\mathrm{helio}}$ \\
   & (J2000) & (J2000) &  & [s] & [$\arcsec$] & [pixel$^{-1}$] & [km s$^{-1}$]\\
  \hline                      
  14853 & 17:44:56.7 & 03:09:39.3 & 2014-08-22 & 7538 & 0.7 & 14/29/80  & $-210.60\pm0.08$\\      
   5659 & 17:44:50.6 & 03:09:09.0 & 2014-08-20 & 4782 & 0.6 & 15/32/84  & $-212.20\pm0.07$\\
    736 & 17:44:51.6 & 03:10:05.8 & 2014-08-20 & 1800 & 0.7 & 5/39/109  & $-213.40\pm0.07$\\
   2314 & 17:44:50.7 & 03:09:57.6 & 2015-08-17 & 5400 & 0.9 & 10/29/88  & $-212.70\pm0.09$\\
  \hline
  \end{tabular}
  \tablefoot{
  \tablefoottext{a}{Shortened versions of the 2MASS catalog identifiers.}  \tablefoottext{b}{From the S/N spectra at wavelengths 4000 {\AA}, 5450 {\AA} and 7500 {\AA}, respectively.}
  }
  \end{table*}
  
  \subsection{Radial velocities}
  Radial velocities (RVs) of our targets were derived by cross correlating a portion of the spectra around H$\alpha$ against a synthetic template spectrum with stellar parameters similar to the ones for our red giant targets (Sect. \ref{Sec:pars}). All four targets have very similar heliocentric velocities, with a mean of $v_{\rm helio}=-212.2\pm0.5$ km s$^{-1}$ and a 1$\sigma$-dispersion of 1.0$\pm$0.4 km s$^{-1}$, this low value being in line with the cluster's moderately low mass and luminosity \citep[$M_V=-6.67$;][]{Harris96}. While the Harris catalog lists a considerably higher velocity of $-162$ km s$^{-1}$, the more recent work of \citet{Dias15} led to  a systemic velocity of $-242\pm$11 km s$^{-1}$ based on five measurements at low resolution ($R\sim2000$). 
  
\section{Stellar atmospheric parameters and abundance analysis}\label{Sec:abanalysis}
Most of our analysis relied on the technique of measuring equivalent widths ($EW$s), the computation of synthetic line profiles and their $EW$s for the respective transition under the assumption of local thermodynamic equilibrium (LTE), and the comparison of both in order to derive elemental abundances for a given set of stellar parameters. For this purpose we employed the June 2014 release of the LTE radiation transfer code MOOG \citep{Sneden73}. For the elements Na and K we provide non-LTE (NLTE) corrections using the respective LTE abundances as a baseline. Throughout this work we have made use of the $\alpha$-enhanced \mbox{ATLAS9} model atmospheres and AODFNEW opacity distribution functions as described in \cite{Castelli04}. Because of reasons explained in Sect. \ref{Sec:pars} we derived the stellar parameters from differential Fe abundances with respect to the benchmark giant star \object{HD~122563}. A MIKE spectrum for this star was kindly provided by I. U. Roederer (private communication).

  \subsection{Atomic data}
  Our line list comprises the line data compiled in \citet{Koch16} for a set of stars with similar parameters and metallicities as well as minor additions from the VALD\footnote{http://vald.astro.uu.se/} database. Lines that seemed to be affected by heavy blending (where the $FWHM$ appeared exceptionally large or were we detected asymmetries in the profile), were rejected as well as those where telluric contamination was present. We accounted for hyperfine splitting of the transitions of the elements Sc \citep{Kurucz95}, V  \citep{Kurucz95}, Mn \citep{denHartog11}, Co \citep{Kurucz95}, Sr \citep{Bergemann12,Hansen13}, Ba \citep{McWilliam98}, La \citep{Lawler01} and Eu \citep{Lawler01b} and assumed a solar scaled isotope distribution for each chemical species \citep{Lodders03}.
  
  \subsection{Equivalent width measurements: EWCODE}
  In order to have a fast and reproducible method of measuring $EW$s without losing the ability to visually inspect every line, we developed a semi-automated Python 3 script we call EWCODE. This code works on RV-corrected, one-dimensional input spectra and fits Gaussian profiles at wavelengths predefined by an input line list. The fitting procedure does not rely on a global continuum or normalization, but on a more local approach of masking the region of the expected feature position. The local continuum is subsequently approximated by fitting a straight line to the flux levels of the neighboring wavelength ranges. Typical widths of these ranges, which decisively determine the stability of the continuum, are two profile $FWHM$s each. This value can be manually adapted. After the continuum placement, a Gaussian profile with the amplitude $a$, the position, and the $FWHM$ as free parameters is fitted via a Levenberg-Marquardt least square algorithm. The whole procedure is iterated with the width of the fitted profile as input mask parameter for the next continuum placement step. The code also provides an estimate for the total $EW$ error, which is computed via
  \begin{equation}
   \sigma_{EW\mathrm{,tot}} = \sqrt{\left(\frac{\partial EW}{\partial a}\Delta a\right)^2 + \left(\frac{\partial EW}{\partial FWHM}\Delta FWHM\right)^2},
  \end{equation}
  following standard Gaussian error propagation. Covariances between $\Delta a$ and $\Delta FWHM$ are not taken into account here, which renders $\sigma_{EW\mathrm{,tot}}$ a conservative upper estimate. One main advantage of this treatment is that it is applicable down to S/N values as low as approximately five, provided that the continuum windows are free of absorption lines \citep[see, e.g.,][]{Fulbright06}. Our metal-poor spectra satisfy this condition over a wide spectral range, that is, there are many isolated transitions without substantial blending above wavelengths of 4000 {\AA} which were reliably fitted by EWCODE. 
  
  We tested the code extensively and found it to be very robust. In particular, we synthesized a fictitious \ion{Fe}{i} line ($\lambda=5000$ {\AA}, $\chi_{\mathrm{ex}}=3$ eV, $\log{gf}=-1$) $10^4$ times for a fixed representative atmosphere ($T_{\mathrm{eff}}=4200$, $\log{g}=1$, [Fe/H]$=-2$ dex, $v_{\mathrm{mic}}=2$ km s$^{-1}$) and an $EW$ interval of [1 m{\AA}, 200 m\AA]. The pixel size was chosen to be $\delta \lambda = 0.02$  {\AA}, representative for the typical values of our MIKE setup. We mimicked instrumental broadening by convolving the spectra with a Gaussian function of $FWHM=0.15$~\AA\ and added noise drawn from Poissonian distributions to implement different S/N characteristics. These profiles were then measured with EWCODE. 
  \begin{figure}
  \centering
  \resizebox{\hsize}{!}{\includegraphics{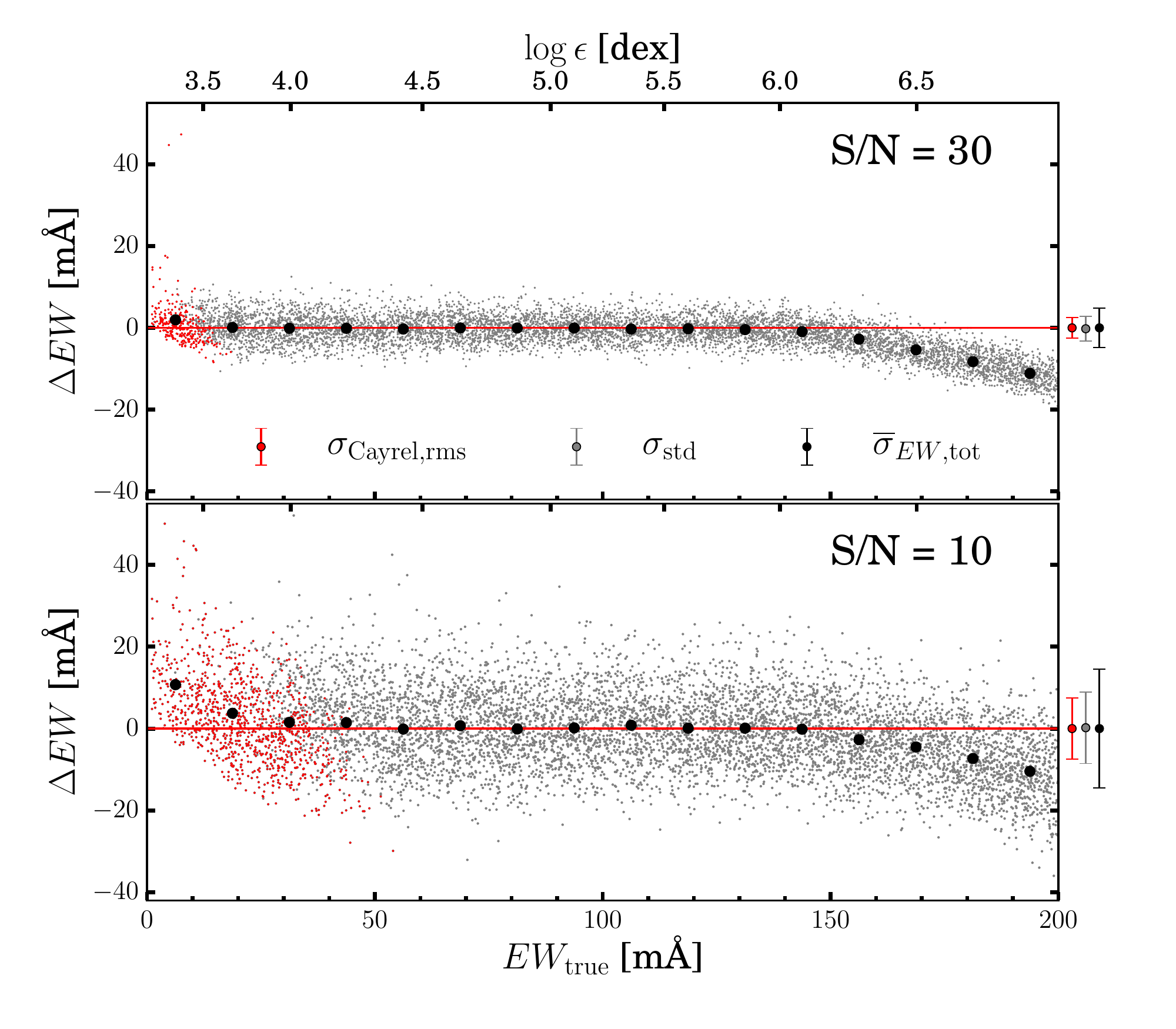}}
     \caption{Residuals of measured and true equivalent widths $\Delta EW$ for S/N of 30 pixel $^{-1}$ (top panel) and 10 pixel $^{-1}$ (lower panel). Measurements which were already rejected within EWCODE are shown as red dots. Thick black points indicate the means of the respective $EW$ bins. Red, gray and black error bars illustrate the theoretical and true rms scatters within the trust region (see text) and the error provided by EWCODE, respectively. The corresponding abundance for this particular synthesized transition (see text) for a given $EW_\mathrm{true}$ is indicated by the top axis.
              }
     \label{Fig:EWCODE}
  \end{figure}
  Figure \ref{Fig:EWCODE} illustrates the residuals of measured and input $EW$s versus the input $EW$s and the resulting logarithmic abundance $\log{\epsilon}$ for S/N$ = 30$ pixel$^{-1}$ and S/N$ = 10$ pixel$^{-1}$. Binned mean deviations are shown in the figure, which indicate that there are two turnoffs from the zero-relation present at the lower (below $15/40$ m{\AA} for S/N$=30/10$) and the upper $EW$ ranges (above 150 m{\AA}), respectively. The former was already rejected within the code, since we only claimed ``detection'' of a feature, if the computed $EW$ is larger than 2.5 times its corresponding individual error $\sigma_{EW\mathrm{,tot}}$ (rejected points are highlighted in the figure). The latter trend sets in, once the assumption of absorption profiles with Gaussian shape looses justification. This case occurs once the core of the line saturates and an increasing amount of absorption moves in the damping wings. Such profiles cannot be correctly fit by EWCODE, because a substantial amount of line depth is not included in the Gaussian fit and therefore the derived $EW$ drops with respect to the true one. A more accurate treatment of these profiles should involve the fitting of Voigt profiles, which can account for Lorentzian wings. We will implement this option in a future version of EWCODE. Here, we have limited our analysis to profiles which can be readily described by our Gaussian approximation.
  
  We distinguish three properties to evaluate the quality of our EWCODE results. First, we introduce the theoretical root mean square (rms) deviation
  \begin{equation}
   \sigma_{\mathrm{Cayrel,rms}} = \frac{1}{S/N}\sqrt{1.5\cdot FWHM\cdot \delta\lambda}   
  ,\end{equation}
  as derived by \citet{Cayrel88} and slightly modified by \citet{Battaglia08}. This relation establishes the lower limit of the measurement uncertainty due to sampling for very high S/N spectra. In fact, EWCODE nearly reaches this limit, judging from the standard deviation $\sigma_{std}$ of points within the trusted region below 150 m{\AA}. For completeness we included the mean of the individual errors $\sigma_{std}=$ $\overline{\sigma}_{EW\mathrm{,tot}}$ provided by the code. At an S/N of $30/10$ pixel$^{-1}$ we reached precisions of $\sigma_{std}=3.0/8.7$ m{\AA} and $\overline{\sigma}_{EW\mathrm{,tot}}=4.8/14.4$ m{\AA}, compared to $\sigma_{\mathrm{Cayrel,rms}}=2.5/7.5$ m{\AA}.
  
  Since many features were to be found in more than one spectral order of the spectrograph, we selected those $EW$s that have been measured in the orders with the highest flux levels. Our $EW$ results are presented in Table \ref{Table:EW_TABLE}.
  
  \begin{table*}
  \caption{Line List}          
  \label{Table:EW_TABLE} 
  \centering                      
  \begin{tabular}{ccccr@{ }c@{ }rr@{ }c@{ }rr@{ }c@{ }rr@{ }c@{ }r} 
  \hline\hline          
  \multicolumn{4}{c}{\phantom{empty}} & \multicolumn{12}{c}{$EW$\tablefootmark{a} [m\AA]}\\
  \cline{5-16}
  $\lambda$ [\AA]& Species & $\chi_{\mathrm{ex}}$ [eV] & $\log{gf}$ & \multicolumn{3}{c}{14853} & \multicolumn{3}{c}{5659} & \multicolumn{3}{c}{736} & \multicolumn{3}{c}{2314}\\
  \hline 
  5688.200 & \ion{Na}{i} & 2.100 & -0.404 &       & ...    &     & 19.0  & $\pm$ & 7.0 &       & ...   &      &       & ...   &      \\
  4571.100 & \ion{Mg}{i} & 0.000 & -5.623 & 105.0 & $\pm$  & 8.0 & 113.0 & $\pm$ & 6.0 & 116.0 & $\pm$ & 12.0 & 133.0 & $\pm$ & 10.0 \\
  4702.990 & \ion{Mg}{i} & 4.350 & -0.440 & 89.0  & $\pm$  & 6.0 & 85.0  & $\pm$ & 6.0 & 92.0  & $\pm$ & 10.0 & 90.0  & $\pm$ & 8.0  \\
  5528.420 & \ion{Mg}{i} & 4.350 & -0.481 & 99.0  & $\pm$  & 7.0 & 98.0  & $\pm$ & 6.0 & 107.0 & $\pm$ & 5.0  & 114.0 & $\pm$ & 9.0  \\
  5711.090 & \ion{Mg}{i} & 4.330 & -1.728 & 16.0  & $\pm$  & 4.0 & 17.0  & $\pm$ & 6.0 & 23.0  & $\pm$ & 4.0  & 9.0   & $\pm$ & 6.0  \\
  5948.540 & \ion{Si}{i} & 5.080 & -1.130 & 17.0  & $\pm$  & 6.0 & 13.0  & $\pm$ & 4.0 & 11.0  & $\pm$ & 4.0  & 19.0  & $\pm$ & 5.0  \\
  7405.770 & \ion{Si}{i} & 5.610 & -0.820 &       & ...    &     & 14.0  & $\pm$ & 4.0 & 10.0  & $\pm$ & 3.0  &       & ...   &      \\
  \hline                                 
  \end{tabular}
  \tablefoot{\tablefoottext{a}{$EW$ uncertainties are the $\sigma_{EW,\mathrm{tot}}$ as provided by EWCODE.} This table  is available in its entirety in electronic form via the CDS.
  }    
  \end{table*}
  
  \subsection{Atmospheric parameters}
  \label{Sec:pars}
  For each stellar model atmosphere the input quantities effective temperature $T_{\mathrm{eff}}$, metallicity [Fe/H] and microturbulence $v_{\mathrm{mic}}$ were derived spectroscopically from line-by-line abundances of the neutral ($\log{\epsilon(\ion{Fe}{i})}$) and singly ionized ($\log{\epsilon(\ion{Fe}{ii})}$) iron species with respect to the abundances estimated for the same lines in the reference star (the choice of lines is further discussed in Sect. \ref{Sec:mic}). Through this treatment we strongly reduced the influences of uncertain oscillator strengths on the atmosphere parameters. Here we adopted $T_{\mathrm{eff}}=4587$ K, $\log{g}=1.61$ dex and [Fe/H$]=-2.64$ dex for HD~122563 as listed by \cite{Heiter15}. For the microturbulence we derived $v_{\mathrm{mic}}=2.16$ km s$^{-1}$ from our spectrum. These parameters agree reasonably well with the expected ones of our stars ($\Delta T_\mathrm{eff} \sim +40$ - $+270$~K, $\Delta \log{g} \sim +0.5$ - $+0.9$ dex, $\Delta$ [Fe/H$] \sim -0.2$ - $-0.3$ dex, $\Delta v_\mathrm{mic} \sim 0$ - $+0.35$ km s$^{-1}$), which makes HD~122563 a reliable benchmark. We note, however, that this star is strongly enhanced in all r-process elements so that an  analysis of the heavy elements in our targets, differentially to HD~122563, would complicate our interpretations. This will be addressed further in Sect. \ref{Sec:indiv_elements}.
  
  The metallicity for HD~122563 stated by \citet{Heiter15} is based on \citet{Jofre14}, who derived [Fe/H] from \ion{Fe}{i} lines which have been corrected for NLTE effects. Hence, as \citet{Jofre14} already pointed out, in order to compare their Fe abundances for HD~122563 to ours, one has to subtract their mean NLTE correction of $\Delta\log{\epsilon(\ion{Fe}{i})}_\mathrm{NLTE}=+0.10$ dex from the reference to get the mean LTE abundance of neutral lines. Likewise, the ionization imbalance of $-0.19$ dex provided in their work has to be subtracted from the reference to end up with [\ion{Fe}{ii}/H]. In addition, since we relied on the solar reference values of \citet{Asplund09} ($\log\epsilon{(\mathrm{Fe})}_\sun=7.50$ dex), one has to bear in mind the discrepancy of 0.05~dex to the \citet{Grevesse07} value ($\log\epsilon{(\mathrm{Fe})}_\sun=7.45$ dex) which was used by \citet{Jofre14}. From our line list and $EW$s for HD~122563, we computed mean absolute abundances of $\log{\epsilon(\ion{Fe}{i})}=4.66\pm0.11$ dex ($1\sigma$ scatter) and $\log{\epsilon(\ion{Fe}{ii})}=4.97\pm0.10$ dex, respectively. These are consistent with the reference value after applying the aforementioned adjustments, that is, $\log{\epsilon(\ion{Fe}{i})}_\mathrm{ref}=4.71$ dex and $\log{\epsilon(\ion{Fe}{ii})}_\mathrm{ref}=5.00$ dex. 
  
  In order to achieve accurate results for the model parameters of our stars, we tuned the atmospheric parameters to simultaneously attain balance between differential \ion{Fe}{i} abundances at low and high excitation potentials ($\chi_{\mathrm{ex}}$) and a zero-trend of abundance with reduced $EW$ ($\log{EW/\lambda}$). We accounted for spectral noise by the propagation of $EW$ errors throughout the process of fitting the respective trends. Therefore we approximated the local curve of growth (COG) for a given $EW$ and a fixed set of transition parameters by calling MOOG twice, with $EW + \sigma_{EW\mathrm{,tot}}$ and $EW - \sigma_{EW\mathrm{,tot}}$. The individual assumed symmetric abundance error was then considered as the average. This value was incorporated as individual line weight for all fit procedures explained below. The methods of establishing the equilibria are not completely independent \citep[see, e.g.,][]{McWilliam95}. Thus, for each set of model parameters we computed the full gradients of the $\log{\epsilon}$-$\chi_{\mathrm{ex}}$- and $\log{\epsilon}$-$\log{EW/\lambda}$ slopes with respect to $T_{\mathrm{eff}}$ and $v_{\mathrm{mic}}$. These were used to iteratively alter the stellar parameters until simultaneous convergence toward zero-slopes was reached. Our final photometric and spectroscopic solutions are listed in Table \ref{Table:ATM_PAR}.
  \begin{table*}
  \caption{Photometric and spectroscopic properties}          
  \label{Table:ATM_PAR}      
  \centering                        
  \begin{tabular}{l@{}c c c c@{}c c c c c}      
  \hline\hline                
  Star-ID & $K_{\mathrm{s,}0}$ & $A_{K\mathrm{s}}$\tablefootmark{a} & $(J-K_\mathrm{s})_0$ & $E(J-K_\mathrm{s})$\tablefootmark{a} & $T_{\mathrm{eff,phot}}$\tablefootmark{b} & $T_{\mathrm{eff,spec}}$ & $\log{g}_{\mathrm{phot}}$ & [Fe/H]\tablefootmark{c}   & $v_{\mathrm{mic}}$ \\
          & [mag] & [mag] & [mag] & [mag] & [K] & [K] & [cm s$^{-2}$] & [dex] & [km s$^{-1}$]\\
  \hline                                                                                                                                  
  14853 & $12.195\pm0.026$ & 0.107 & 0.658  & 0.141 & $4680\pm130$ & $4542\pm42$ & $1.14\pm0.09$ & $-2.41\pm0.05$ & $2.32\pm0.14$ \\      
   5659 & $11.948\pm0.028$ & 0.106 & 0.702  & 0.141 & $4530\pm135$ & $4415\pm33$ & $1.03\pm0.09$ & $-2.33\pm0.04$ & $2.16\pm0.14$ \\
    736 & $11.495\pm0.023$ & 0.109 & 0.714  & 0.144 & $4499\pm107$ & $4354\pm33$ & $0.84\pm0.09$ & $-2.40\pm0.04$ & $2.25\pm0.14$ \\
   2314 & $11.208\pm0.024$ & 0.109 & 0.750  & 0.144 & $4384\pm109$ & $4316\pm36$ & $0.72\pm0.09$ & $-2.32\pm0.05$ & $2.51\pm0.14$ \\
  \hline                                  
  \end{tabular}
  \tablefoot{
  \tablefoottext{a}{Extracted from the extinction maps of \citet{Schlafly11}}
  \tablefoottext{b}{Derived from the $T_\mathrm{eff, phot}$-$(J-K_\mathrm{s})_0$-[Fe/H] relations of \citet{Casagrande10}.}
  \tablefoottext{c}{As computed from \ion{Fe}{ii} abundances. Errors are based on $\sigma_{\log{\epsilon}, EW}$ only.}}
  \end{table*}
  
    \subsubsection{Effective temperature}\label{Sec:Teff}
\begin{figure}
    \centering
    \resizebox{\hsize}{!}{\includegraphics{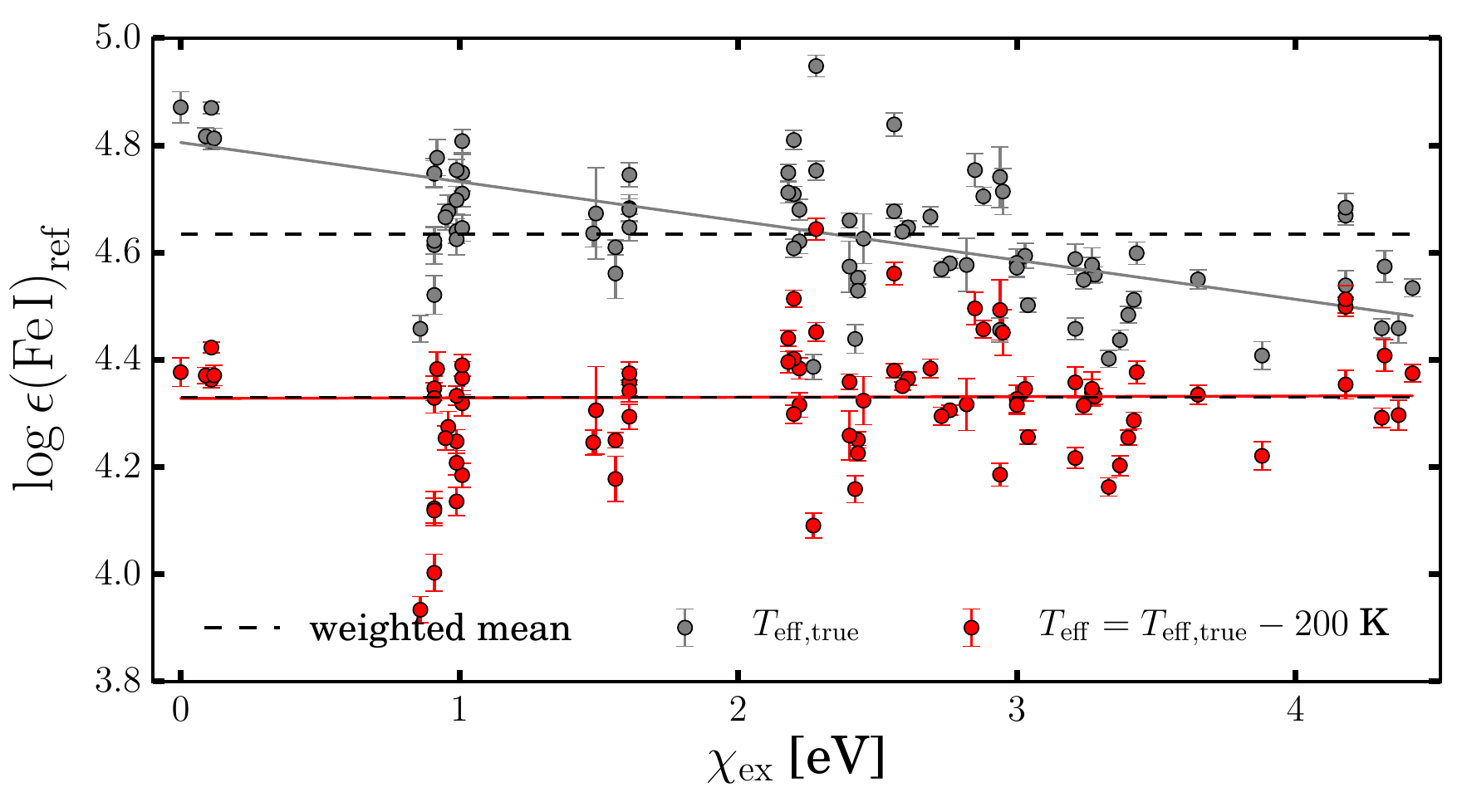}}
       \caption{Absolute iron abundance trends with excitation potential $\chi_\mathrm{ex}$ of \ion{Fe}{i} lines in HD~122563 for the reference effective temperature ($T_\mathrm{eff, true}=4587$ K, gray points) and the one that satisfies excitation equilibrium best ($T_\mathrm{eff}=4387$ K, red points). Error bars are the results of propagating $\sigma_{EW\mathrm{, tot}}$ through the abundance analysis. The weighted means for each realization are represented by the black dashed lines.
       }
       \label{Fig:NLTE_Teff}
    \end{figure}

    The criterion that most affects the choice of $T_{\mathrm{eff}}$  is the abundance trend of neutral lines with low excitation potential $\chi_{\mathrm{ex}}$. To emphasize the importance of a differential treatment, we note the implications indicated by Fig. \ref{Fig:NLTE_Teff}, where we present the $\log{\epsilon(\ion{Fe}{i})}$-$\chi_{\mathrm{ex}}$ relations of HD~122563 as fitted by a least-squares method to the absolute abundances derived from our line list and $EW$-measurements.
    We realized two temperatures while keeping the other parameters fixed at their reference values. The data points at higher abundances were computed using the reference of $T_{\mathrm{eff,true}}=4587$ K. The fitted linear trend of $-0.07$ dex eV$^{-1}$ points to a severe excitation imbalance, which could be resolved by lowering $T_{\mathrm{eff}}$ by 200~K. This, however, would represent an unfeasibly low temperature with respect to the value of \citet{Heiter15}, which should be very accurate considering its origin from bolometric flux calibrations. In addition, this lower temperature would lead to a higher theoretical occupation number of \ion{Fe}{i} low excitation states and thus strengthen their lines. This, in turn, leads to a lower derived abundance for a fixed $EW$. The effect on the weighted average of \ion{Fe}{i} abundances for HD~122563 turns out to be as large as $\Delta(\log{\epsilon(\ion{Fe}{i})})=-0.3$ dex. \citet{Mashonkina11} also encountered a negative excitation trend for this star. They were able to considerably flatten their slope from $ -0.054$ dex eV$^{-1}$ to $-0.030$ dex eV$^{-1}$ by introducing NLTE effects in their line calculations but still could not completely get rid of the imbalance. We note here that there are possible additional sources for excitation imbalances. These comprise, for example, systematic errors in the oscillator strengths $\log{gf}$, 3D- and hydrodynamical effects as well as systematic errors in the model atmospheres. Because of the similarity of our stars to HD~122563, we expected them to be subject to the same slope offset at the ``true'' $T_\mathrm{eff}$ if we used absolute abundances. This could lead to an underestimated $T_\mathrm{eff}$ by 200~K and consequently a lower $\log{\epsilon(\ion{Fe}{i})}$. Thus we determined our stellar parameters from a strict line-by-line differential analysis relative to the respective \ion{Fe}{i} and \ion{Fe}{ii} abundances established by HD~122563, when employing the \citet{Heiter15} value of $T_\mathrm{eff, true} = 4587$~K. 
    
\begin{figure}
    \centering
    \resizebox{\hsize}{!}{\includegraphics{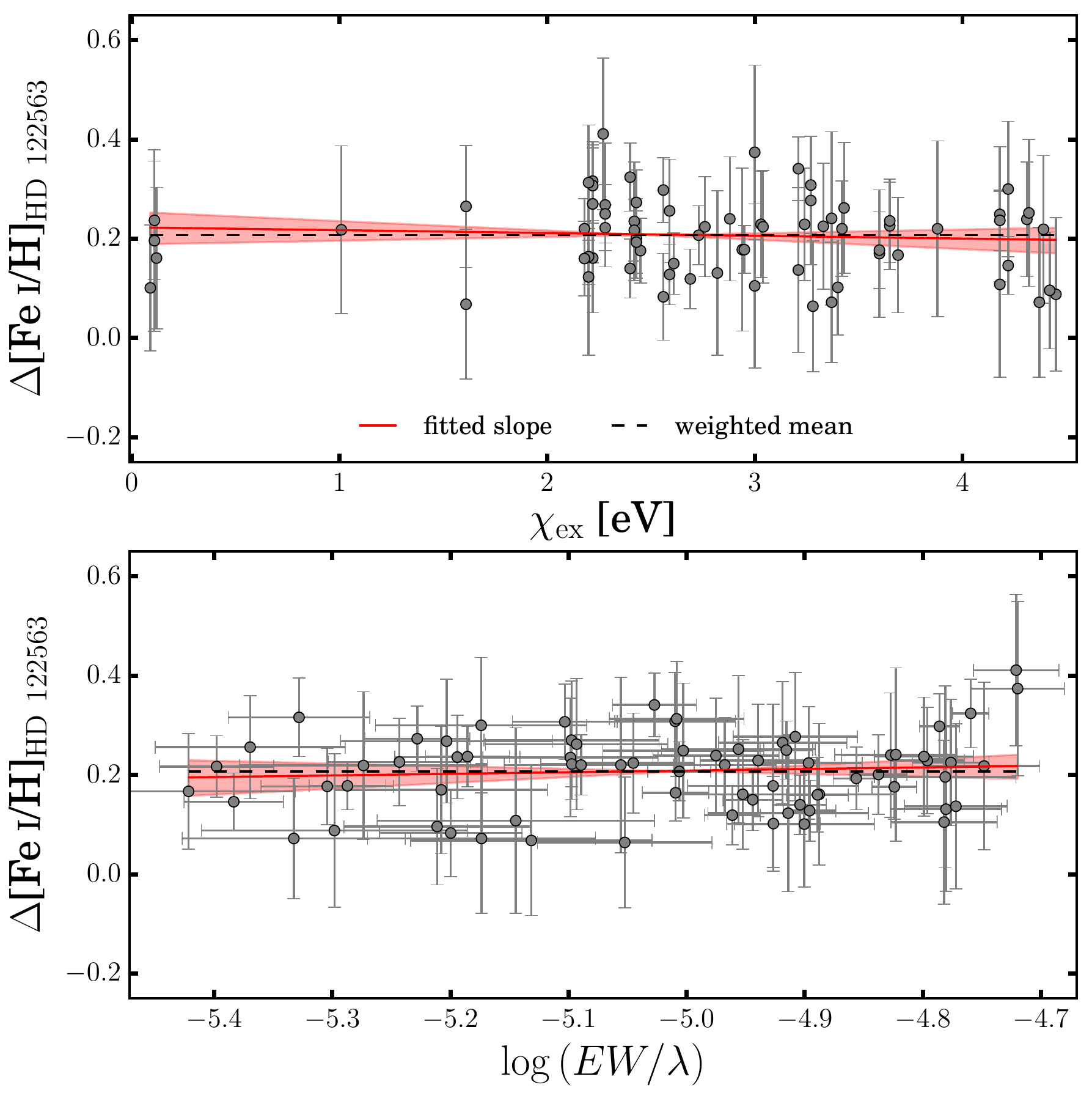}}
      \caption{Exemplary final solutions for $T_\mathrm{eff, spec}$ and $v_\mathrm{mic}$ from balanced differential \ion{Fe}{i} abundances with excitation potential (top) and line strength (bottom) for the star 5659. Errors are again computed from the $EW$ uncertainties. The red shaded areas emphasize the slope errors corresponding to $\pm33$ K and $\pm0.14$ km s$^{-1}$.
              }
      \label{Fig:ATMOS_DEMO}
    \end{figure}

    In doing so we removed another issue, that becomes evident in Fig. \ref{Fig:NLTE_Teff}. Apparently, the abundance scatter is not well reproduced by the individual errors due to $\sigma_{EW\mathrm{,tot}}$, only. Since this is already an upper estimate, the substantially larger scatter was believed to originate from the systematic uncertainty of the line properties, especially of $\log{gf}$. Following the differential approach, in our program stars we could in fact reduce the scatter, which is now significantly better described by $\sigma_{EW\mathrm{,tot}}$ only. This can be seen in the top panel of Fig. \ref{Fig:ATMOS_DEMO}, where we show the findings for the star 5659. For this star the scatter in $\log{\epsilon(\ion{Fe}{i})}$ decreased from 0.15 dex for absolute abundances to 0.08 dex in the differential case. 
    We clipped those points from the fit that deviated by more than 2.5 times their corresponding uncertainty. These might just represent features with additional blending which was not recognized by EWCODE or profile fits with underestimated EW uncertainties. Under the above assumption we computed a $1\sigma$ slope error (indicated in Fig. \ref{Fig:ATMOS_DEMO} corresponding to $\pm0.013$ dex eV$^{-1}$), which was used in the code to estimate the respective temperature uncertainties of our sample between 33~K and 42~K. At this point, we note that these are differential uncertainties with respect to the HD~122563 reference, which according to \citet{Heiter15} has an uncertainty of 60~K.
    
    For completeness, we included the photometric $T_\mathrm{eff}$ using the relations of \citet{Casagrande10} based on our dereddened $(J-K_\mathrm{s})_0$ colors. We note that they are on average 120 K hotter than their spectroscopic counterparts. However, we emphasize the large uncertainties introduced by the photometry errors. In principle, it would be desirable to compute the more robust temperatures from $V-K_\mathrm{s}$ colors, but --as mentioned previously-- we don't have reliable $V$ magnitudes for our stars at hand. Using $V-K_\mathrm{s}$ colors nevertheless -- incorporating the \citet{Dotter11} V magnitudes-- would increase the listed $T_\mathrm{eff}$ by $\sim 300$ K on average. Moreover, the extinction maps by \citet{Schlafly11} are badly resolved for our purposes, that is, if any differential reddening \citep[see, e.g.,][]{Hendricks12} was present, the $T_\mathrm{eff, phot}$ of our stars would be grossly misjudged. For NGC~6426, \citet{Dotter11} reported ``signs of differential reddening''. Therefore, we claim our spectroscopic, reddening-independent results to be more reliable.

    \subsubsection{Surface gravity and metallicity}\label{Sec:logg}
    The surface gravity entering our models was calculated during each iteration step from the relation
    \begin{equation}\label{Eq:LOGG}
     \log{g} = \log{g_\sun} + \log{\frac{M}{M_\sun}} + 4 \log{\frac{T_{\mathrm{eff}}}{T_{\mathrm{eff,\sun}}}} - 0.4\left( M_{\mathrm{bol,\sun}} - M_K - BC_K\right),
    \end{equation}
    where we used  $\log{g_\sun}= 4.44$, $T_{\mathrm{eff,\sun}} = 5777$ K, $M_{\mathrm{bol,\sun}} = 4.74$ and a stellar mass of $M=0.8M_\sun$ for our red giants \citep[see, e.g.,][]{Koch08a}. We computed the absolute magnitude $M_K$ using an adopted distance of 20.6 kpc \citep{Harris96}. In addition, we employed the relation provided in equation (4) of \citet{Buzzoni10} for the bolometric K$_\mathrm{s}$-band correction $BC_K$. The presented uncertainties of 0.09 dex originate from a Monte-Carlo error propagation of our $T_{\mathrm{eff}}$ and $K_\mathrm{s}$ errors and a mass range of $\pm0.15M_\sun$ \citep[see][]{Roederer16}. \citet{Lind12} showed for metal-poor giant stars that the spectroscopic method of deducing $\log{g}$ (with absolute abundances) is heavily biased by NLTE effects on \ion{Fe}{i} lines. In particular, their NLTE calculations indicate that due to a stronger net UV-radiation field compared to the LTE case the ionization equilibrium gets shifted toward the ionized state, leading to a weakening of neutral Fe lines. Thus, in LTE, one underestimates $\log{\epsilon(\ion{Fe}{i})}$. This is confirmed by the derivation of NLTE abundances for several of our HD~122563 \ion{Fe}{i} lines from our measured $EW$s using the grid of \citet{Bergemann12_NLTE} and \citet{Lind12}\footnote{Data obtained from the INSPECT database, version 1.0 (www.inspect-stars.net)}. We found typical corrections of the order of $\Delta\log{\epsilon(\ion{Fe}{i})}_\mathrm{NLTE}\approx +0.14$~dex for low-excitation lines around $\chi_\mathrm{ex}\approx0$~eV and $\Delta\log{\epsilon(\ion{Fe}{i})}_\mathrm{NLTE}\approx +0.26$~dex at $\chi_\mathrm{ex}\approx4$~eV. Because \ion{Fe}{ii} is already the dominant species, overionization only marginally affects its abundance (typical corrections for HD~122563 range from -0.002 dex to -0.010 dex). A decrease in the model $\log{g}$, however, tends to lower $\log{\epsilon(\ion{Fe}{ii})}$ and has a vanishingly small impact on $\log{\epsilon(\ion{Fe}{i})}$. Consequently, by forcing both abundances into equilibrium, one would underestimate the surface gravity and the metallicity, respectively. We report on a mean ionization imbalance of $\log{\epsilon(\ion{Fe}{ii})}-\log{\epsilon(\ion{Fe}{i})}\sim 0.3$ dex in terms of absolute abundances in the four program stars. Our usage of a differential measurement levels out the mentioned discrepancies to some degree, because we found $\log{\epsilon(\ion{Fe}{ii})}-\log{\epsilon(\ion{Fe}{i})} = 0.31$ dex for HD~122563 and its literature stellar parameters. Hence, our photometric gravities appear to establish ionization balance in the differential case. Nevertheless, we would like to stress that our stars span a range of effective temperatures and gravities below the values of HD~122563 and expose higher $\mathrm{[Fe/H]}$. Owing to the fact that \citet{Lind12} also pointed out that the magnitude of NLTE corrections depends strongly on the particular choice of stellar parameters and Fe lines, we cannot claim to be fully unaffected by them. Accordingly, the metallicities [Fe/H] in this work are based on the weighted mean of the \ion{Fe}{ii} abundances, which we consider more reliable \citep[see also][]{Kraft03}. Again, we applied the clipping of outliers that deviated by more than $2.5 \sigma_{\log{\epsilon}, EW}$. This mean value was computed during each iteration and set the metallicity of the successive step (for a more detailed discussion on Fe abundances see Sect. \ref{Sec:iron}).     
    
    \subsubsection{Microturbulence}\label{Sec:mic}
    Since our microturbulence parameter $v_{\mathrm{mic}}$ was tuned toward erasing the slope of the $\log{\epsilon(\ion{Fe}{i})}$ vs. line strength (reduced $EW$, $\log{EW/\lambda}$) relation, we again had to ensure that this slope was free of intrinsic influences causing the ``true'' microturbulent velocity to produce non-zero slopes. First, we followed the common procedure of limiting the analysis to weak and intermediately strong lines, that is, excluding the strongest, possibly saturated lines, that originate from the flat part of the COG. The reasoning behind this is that these features emerge from the highest, most dilute atmospheric layers of a star, where our LTE treatment and model atmospheres as well as the assumption of Gaussian line profiles are likely to fail. As a consequence we established an upper limit for the iron line strength of $\log{EW/\lambda}=-4.7$. Since the uncertainties of $\log{\epsilon(\ion{Fe}{i})}$ and $\log{EW/\lambda}$ are not independent, but coupled through the COG, a systematic bias for the slope and therefore the microturbulence can be expected. In order to decouple these quantities, \citet{Magain84} suggested using theoretical line strengths when fitting the slope. Unfortunately, this treatment does not account for an additional effect that becomes apparent in Fig. \ref{Fig:EWCODE}. When working with low S/N spectra one encounters a selection bias at the lower end of the $EW$ distribution due to the noise-induced detection threshold. This threshold runs diagonally in the $\Delta EW$-$EW_{\mathrm{true}}$ plane. Hence, low-$EW$ lines are more likely to be over- than underestimated in their magnitude with respect to their true values. Once propagated through the abundance analysis, one will end up with systematically higher $\log{\epsilon(\ion{Fe}{i})}$ in the regime of low line strength and therefore measure a negative slope for the correct $v_{\mathrm{mic}}$. The vast majority of our iron lines resides in Echelle orders with S/N well above 30, so we decided to apply a reasonable lower $EW$ cut of 20 m{\AA}. In his Monte-Carlo analysis with various S/N characteristics, \citet{Mucciarelli11} found that a proper treatment of the errors within the classical approach of finding $v_{\mathrm{mic}}$ is superior to the Magain method. Following the suggestions made in that work, we used an orthogonal distance regression algorithm to fit the desired slopes. This enabled us to treat both abscissa and ordinate uncertainties. As an example we present the fit outcome for star 5659 in the lower panel of Fig. \ref{Fig:ATMOS_DEMO}. Again, we used the slope error to constrain a respective $v_{\mathrm{mic}}$ uncertainty.

\section{Abundance results}\label{Sec:abres}
In Table \ref{Table:AB_RESI} we present the weighted mean abundance ratios using our final model atmosphere solutions. While using differential abundances of Fe transitions to constrain the stellar parameters, we relied on absolute abundances -- $\log{gf}$ values-- for our final elemental abundance results. We emphasize the fact that --depending on $\chi_\mathrm{ex}$-- this treatment potentially introduces systematic errors to the abundances (see Sect. \ref{Sec:Teff}), which ultimately leads to increased values for the scatter around the weighted mean in Table \ref{Table:AB_RESI}. All ratios of neutral and ionized species in the following sections were computed relative to their counterparts \ion{Fe}{i} and \ion{Fe}{ii}. Our results for the mean chemical abundances of NGC~6426 together with the rms scatter of the stars are shown in Table \ref{Table:AB_RES_CLUSTER_MEAN}.

\begin{table*}
\caption{Final abundance results for individual stars.}
\label{Table:AB_RESI}
\centering
\begin{tabular}{@{\extracolsep{6pt}}lcccccrccccr}
\hline\hline
 & &  \multicolumn{5}{c}{14853}  &  \multicolumn{5}{c}{5659}\\
\cline{3-7} \cline{8-12}\\[-2.1ex]
Species X & Z & $\log{\epsilon}$ & [X/Fe]\tablefootmark{a} & $\sigma_{\mathrm{stat}}$\tablefootmark{b} & $\sigma_\mathrm{tot}$ & \multicolumn{1}{c}{$N_{\mathrm{lines}}$} & $\log{\epsilon}$ & [X/Fe]\tablefootmark{a} & $\sigma_{\mathrm{stat}}$\tablefootmark{b} & $\sigma_\mathrm{tot}$ & \multicolumn{1}{c}{$N_{\mathrm{lines}}$}\\ 
\hline 
$[$\ion{Fe}{i}/H] & 26 & 4.85 & \llap{$-$}2.65 & 0.16 & 0.18 & 94 & 4.84 & \llap{$-$}2.66 & 0.15 & 0.16 & 111\\
$[$\ion{Fe}{ii}/H] & 26 & 5.10 & \llap{$-$}2.40 & 0.12 & 0.13 & 9 & 5.17 & \llap{$-$}2.33 & 0.13 & 0.14 & 12\\
\hdashline
\ion{Na}{i} & 11 & ... & ... & ... & ... & ... & 3.94 & 0.36 & 0.20 & 0.30 & 1\\
\ion{Mg}{i} & 12 & 5.44 & 0.49 & 0.06 & 0.09 & 4 & 5.35 & 0.41 & 0.07 & 0.09 & 4\\
\ion{Si}{i} & 14 & 5.27 & 0.41 & 0.26 & 0.26 & 2 & 5.44 & 0.59 & 0.18 & 0.18 & 3\\
\ion{K}{i} & 19 & 2.96 & 0.58 & 0.08 & 0.10 & 1 & 2.91 & 0.54 & 0.05 & 0.07 & 1\\
\ion{Ca}{i} & 20 & 3.93 & 0.24 & 0.23 & 0.24 & 11 & 3.96 & 0.28 & 0.24 & 0.24 & 13\\
\ion{Sc}{ii} & 21 & 0.66 & \llap{$-$}0.09 & 0.17 & 0.17 & 7 & 0.71 & \llap{$-$}0.11 & 0.20 & 0.20 & 9\\
\ion{Ti}{i} & 22 & 2.54 & 0.24 & 0.06 & 0.10 & 9 & 2.42 & 0.13 & 0.12 & 0.14 & 11\\
\ion{Ti}{ii} & 22 & 2.79 & 0.24 & 0.19 & 0.21 & 10 & 2.81 & 0.19 & 0.13 & 0.15 & 12\\
\ion{V}{i} & 23 & 1.05 & \llap{$-$}0.23 & 0.14 & 0.16 & 1 & 1.18 & \llap{$-$}0.09 & 0.24 & 0.25 & 2\\
\ion{Cr}{i} & 24 & 2.76 & \llap{$-$}0.23 & 0.07 & 0.10 & 4 & 2.74 & \llap{$-$}0.24 & 0.02 & 0.08 & 5\\
\ion{Mn}{i} & 25 & 2.33 & \llap{$-$}0.45 & 0.11 & 0.13 & 3 & 2.30 & \llap{$-$}0.47 & 0.34 & 0.34 & 5\\
\ion{Co}{i} & 27 & 2.18 & \llap{$-$}0.16 & 0.45 & 0.46 & 1 & ... & ... & ... & ... & ...\\
\ion{Ni}{i} & 28 & 3.69 & 0.12 & 0.22 & 0.23 & 9 & 3.61 & 0.05 & 0.13 & 0.14 & 11\\
\ion{Zn}{i} & 30 & 2.21 & 0.30 & 0.16 & 0.16 & 1 & 2.32 & 0.42 & 0.13 & 0.13 & 2\\
\ion{Sr}{ii} & 38 & 0.55 & 0.08 & 0.19 & 0.19 & 1 & 0.19 & \llap{$-$}0.35 & 0.28 & 0.28 & 1\\
\ion{Y}{ii} & 39 & \llap{$-$}0.57 & \llap{$-$}0.38 & 0.06 & 0.07 & 2 & \llap{$-$}0.49 & \llap{$-$}0.37 & 0.05 & 0.06 & 5\\
\ion{Zr}{ii} & 40 & 0.46 & 0.28 & 0.01 & 0.04 & 2 & 0.58 & 0.33 & 0.08 & 0.09 & 3\\
\ion{Ba}{ii} & 56 & \llap{$-$}0.32 & \llap{$-$}0.10 & 0.08 & 0.11 & 3 & \llap{$-$}0.31 & \llap{$-$}0.16 & 0.06 & 0.12 & 3\\
\ion{La}{ii} & 57 & \llap{$-$}1.10 & 0.20 & 0.27 & 0.27 & 1 & \llap{$-$}1.30 & \llap{$-$}0.07 & 0.20 & 0.20 & 1\\
\ion{Nd}{ii} & 60 & \llap{$-$}0.41 & 0.57 & 0.00 & 0.04 & 2 & \llap{$-$}0.49 & 0.42 & 0.11 & 0.11 & 2\\
\ion{Sm}{ii} & 62 & ... & ... & ... & ... & ... & \llap{$-$}0.94 & 0.43 & 0.20 & 0.20 & 1\\
\ion{Eu}{ii} & 63 & \llap{$-$}1.46 & 0.42 & 0.13 & 0.13 & 1 & \llap{$-$}1.44 & 0.37 & 0.14 & 0.14 & 1\\
\hline
 & &  \multicolumn{5}{c}{736}  &  \multicolumn{5}{c}{2314}\\
\cline{3-7} \cline{8-12}\\[-2.1ex]
Species X & Z & $\log{\epsilon}$ & [X/Fe]\tablefootmark{a} & $\sigma_{\mathrm{stat}}$\tablefootmark{b} & $\sigma_\mathrm{tot}$ & \multicolumn{1}{c}{$N_{\mathrm{lines}}$} & $\log{\epsilon}$ & [X/Fe]\tablefootmark{a} & $\sigma_{\mathrm{stat}}$\tablefootmark{b} & $\sigma_\mathrm{tot}$ & \multicolumn{1}{c}{$N_{\mathrm{lines}}$}\\ 
\hline 
$[$\ion{Fe}{i}/H] & 26 & 4.88 & \llap{$-$}2.62 & 0.13 & 0.14 & 121 & 4.89 & \llap{$-$}2.61 & 0.14 & 0.16 & 110\\
$[$\ion{Fe}{ii}/H] & 26 & 5.14 & \llap{$-$}2.36 & 0.18 & 0.18 & 14 & 5.23 & \llap{$-$}2.27 & 0.15 & 0.16 & 12\\
\hdashline
\ion{Na}{i} & 11 & ... & ... & ... & ... & ... & ... & ... & ... & ... & ...\\
\ion{Mg}{i} & 12 & 5.47 & 0.49 & 0.16 & 0.17 & 4 & 5.37 & 0.38 & 0.15 & 0.17 & 3\\
\ion{Si}{i} & 14 & 5.32 & 0.43 & 0.12 & 0.12 & 3 & 5.44 & 0.54 & 0.10 & 0.10 & 2\\
\ion{K}{i} & 19 & 2.92 & 0.51 & 0.05 & 0.08 & 1 & 2.90 & 0.48 & 0.08 & 0.10 & 1\\
\ion{Ca}{i} & 20 & 3.98 & 0.26 & 0.20 & 0.21 & 16 & 3.94 & 0.21 & 0.20 & 0.21 & 18\\
\ion{Sc}{ii} & 21 & 0.69 & \llap{$-$}0.10 & 0.33 & 0.33 & 9 & 0.78 & \llap{$-$}0.10 & 0.21 & 0.21 & 7\\
\ion{Ti}{i} & 22 & 2.56 & 0.23 & 0.16 & 0.18 & 15 & 2.55 & 0.21 & 0.11 & 0.15 & 14\\
\ion{Ti}{ii} & 22 & 2.82 & 0.23 & 0.17 & 0.19 & 13 & 2.83 & 0.15 & 0.17 & 0.20 & 13\\
\ion{V}{i} & 23 & 1.06 & \llap{$-$}0.25 & 0.25 & 0.26 & 1 & 1.03 & \llap{$-$}0.29 & 0.13 & 0.16 & 1\\
\ion{Cr}{i} & 24 & 2.81 & \llap{$-$}0.21 & 0.08 & 0.11 & 5 & 2.75 & \llap{$-$}0.28 & 0.09 & 0.13 & 6\\
\ion{Mn}{i} & 25 & 2.39 & \llap{$-$}0.42 & 0.12 & 0.13 & 4 & 2.48 & \llap{$-$}0.34 & 0.20 & 0.21 & 4\\
\ion{Co}{i} & 27 & ... & ... & ... & ... & ... & 2.32 & \llap{$-$}0.06 & 0.38 & 0.40 & 1\\
\ion{Ni}{i} & 28 & 3.64 & 0.04 & 0.16 & 0.17 & 14 & 3.70 & 0.09 & 0.20 & 0.21 & 15\\
\ion{Zn}{i} & 30 & 2.21 & 0.27 & 0.21 & 0.21 & 1 & 2.52 & 0.57 & 0.17 & 0.17 & 1\\
\ion{Sr}{ii} & 38 & 0.45 & \llap{$-$}0.06 & 0.21 & 0.21 & 1 & 0.76 & 0.16 & 0.14 & 0.14 & 1\\
\ion{Y}{ii} & 39 & \llap{$-$}0.47 & \llap{$-$}0.32 & 0.09 & 0.09 & 5 & \llap{$-$}0.30 & \llap{$-$}0.24 & 0.22 & 0.22 & 3\\
\ion{Zr}{ii} & 40 & ... & ... & ... & ... & ... & 0.54 & 0.23 & 0.33 & 0.33 & 1\\
\ion{Ba}{ii} & 56 & \llap{$-$}0.29 & \llap{$-$}0.11 & 0.08 & 0.12 & 3 & \llap{$-$}0.26 & \llap{$-$}0.17 & 0.14 & 0.17 & 3\\
\ion{La}{ii} & 57 & ... & ... & ... & ... & ... & \llap{$-$}1.46 & \llap{$-$}0.29 & 0.28 & 0.28 & 1\\
\ion{Nd}{ii} & 60 & \llap{$-$}0.53 & 0.41 & 0.45 & 0.45 & 2 & \llap{$-$}0.28 & 0.57 & 0.25 & 0.25 & 3\\
\ion{Sm}{ii} & 62 & ... & ... & ... & ... & ... & \llap{$-$}1.04 & 0.27 & 0.15 & 0.15 & 1\\
\ion{Eu}{ii} & 63 & ... & ... & ... & ... & ... & \llap{$-$}1.59 & 0.16 & 0.19 & 0.19 & 1\\
\hline
\end{tabular}
\tablefoot{
\tablefoottext{a} {The ratios [X/Fe] were computed relative to the abundances of the corresponding iron species of the same ionization state.}
\tablefoottext{b} {For $N_\mathrm{lines}>1$ statistical errors are based on the rms scatter of the involved lines, while for $N_\mathrm{lines}=1$ $\sigma_{\log{\epsilon},EW}$ is stated.}
}
\end{table*}

\begin{table}
\caption{Mean abundances for NGC~6426}
\label{Table:AB_RES_CLUSTER_MEAN}
\centering
\begin{tabular}{@{\extracolsep{6pt}}lc@{}c@{}c@{}c}
\hline\hline
Species X    & Z & $\langle\mathrm{[X/Fe]}\rangle$\tablefootmark{a} & $\sigma_{\mathrm{[X/Fe],rms}}$ & $N_\mathrm{stars}$ \\
\hline
$[$\ion{Fe}{i}/H] & 26 &\llap{$-$}2.64 & 0.02 & 4   \\
$[$\ion{Fe}{ii}/H] & 26 & \llap{$-$}2.34 & 0.05 & 4 \\
\hdashline                                        
\ion{Na}{i} & 11 & 0.36 & ... & 1                \\
\ion{Mg}{i} & 12 & 0.44 & 0.05 & 4                \\
\ion{Si}{i} & 14 & 0.49 & 0.07 & 4                \\
\ion{K}{i} & 19 & 0.53 & 0.04 & 4                 \\
\ion{Ca}{i} & 20 & 0.25 & 0.03 & 4                \\
\ion{Sc}{ii} & 21 & \llap{$-$}0.10 & 0.01 & 4     \\
\ion{Ti}{i} & 22 & 0.20 & 0.04 & 4                \\
\ion{Ti}{ii} & 22 & 0.20 & 0.04 & 4               \\
\ion{V}{i} & 23 & \llap{$-$}0.21 & 0.08 & 4       \\
\ion{Cr}{i} & 24 & \llap{$-$}0.24 & 0.03 & 4      \\
\ion{Mn}{i} & 25 & \llap{$-$}0.42 & 0.05 & 4      \\
\ion{Co}{i} & 27 & \llap{$-$}0.11 & 0.05 & 2      \\
\ion{Ni}{i} & 28 & 0.08 & 0.03 & 4                \\
\ion{Zn}{i} & 30 & 0.39 & 0.12 & 4                \\
\ion{Sr}{ii} & 38 & \llap{$-$}0.04 & 0.19 & 4     \\
\ion{Y}{ii} & 39 & \llap{$-$}0.33 & 0.06 & 4      \\
\ion{Zr}{ii} & 40 & 0.28 & 0.04 & 3               \\
\ion{Ba}{ii} & 56 & \llap{$-$}0.14 & 0.03 & 4     \\
\ion{La}{ii} & 57 & \llap{$-$}0.05 & 0.20 & 3     \\
\ion{Nd}{ii} & 60 & 0.49 & 0.08 & 4               \\
\ion{Sm}{ii} & 62 & 0.35 & 0.08 & 2               \\
\ion{Eu}{ii} & 63 & 0.32 & 0.11 & 3               \\
\hline
\end{tabular}
\tablefoot{
\tablefoottext{a} {Ratios of neutral and ionized species with respect to iron are computed relative to the abundances of the corresponding iron species of the same ionization state.}}
\end{table}
 
  \subsection{Abundance errors}
  In order to quantify the errors of our abundance analysis with respect to the uncertainties of the stellar parameters, we followed the procedure described in \citet{Koch14}. In particular, we repeated the analysis for all the stars with the stellar parameters being lowered and elevated by the uncertainties given in Table \ref{Table:ATM_PAR}, while keeping the others fixed. To account for uncertainties of the opacity distribution function with respect to the $\alpha$-content, we also ran a computation with ODFNEW opacity distributions resembling solar [$\alpha$/Fe] ratios. 
  
  In Table \ref{Table:ABUND_SYST_ERR} we demonstrate the impact of stellar parameters on the abundances and how their uncertainties propagate into the corresponding abundance uncertainties. This is shown for the stars 14853 and 2314 which have the largest separation in parameter space and should therefore be representative for the whole sample. The total uncertainty due to the model atmospheres used, $\sigma_\mathrm{atm}$, was computed by quadratically summing all the mean absolute deviations. For this calculation we divided the ODFNEW contribution by four in order to mimic 0.1 dex of $\alpha$-content uncertainty. We emphasize that we did not account for correlations of the parameters such as the $T_\mathrm{eff}$-$\log{g}$ dependency \citep[see][for further discussion]{McWilliam95}. Hence, the given uncertainties are to be seen as upper estimates. We find variations of $T_\mathrm{eff}$ to have the largest impact, which is especially pronounced for low excitation transitions of neutral species. Except in a few cases, the $\sigma_\mathrm{atm}$ reside well below 0.1 dex.
  \begin{table*}
  \renewcommand{\arraystretch}{1.5}
  \caption{Abundance deviations from uncertainties in the stellar parameters for the stars 14853 and 2314.}          
  \label{Table:ABUND_SYST_ERR}      
  \centering  
  \resizebox{\textwidth}{!}{%
  \begin{tabular}{@{\extracolsep{6pt}}lc@{}c@{}c@{}c@{}c@{}cc@{}c@{}c@{}c@{}c@{}c}
  \hline\hline
  & \multicolumn{6}{c}{14853} & \multicolumn{6}{c}{2314}\\
  \cline{2-7} \cline{8-13}
  & $\Delta T_{\rm eff}$ & $\Delta\,\log\,g$ & $\Delta$[M/H] & $\Delta v_\mathrm{mic}$ & & & $\Delta T_{\rm eff}$ & $\Delta\,\log\,g$ &  $\Delta$[M/H] & $\Delta v_\mathrm{mic}$ & &  \\

  \raisebox{1.5ex}[-1.5ex]{Species}  & $\pm$42\,K  & $\pm$0.09\,dex  & $\pm$0.05\,dex & $\pm$0.14\,km\,s$^{-1}$ & \raisebox{1.5ex}[-1.5ex]{ODF} & \raisebox{1.5ex}[-1.5ex]{$\sigma_{\rm atm}$} & $\pm$36\,K  & $\pm$0.09\,dex & $\pm$0.05\,dex & $\pm$0.14\,km\,s$^{-1}$ & \raisebox{1.5ex}[-1.5ex]{ODF} & \raisebox{1.5ex}[-1.5ex]{$\sigma_{\rm atm}$} \\
  \hline
  \ion{Mg}{i} & $^{+0.04}_{-0.06}$ & $^{-0.02}_{+0.01}$ & $^{-0.02}_{+0.01}$ & $^{-0.04}_{+0.03}$&$+$0.04&0.07 & $\pm$0.05 & $\mp$0.02 & $^{-0.01}_{+0.02}$ & $^{-0.04}_{+0.05}$&$+$0.06&0.07\\
  \ion{Si}{i} & $\pm$0.02 & <0.01 & <0.01 & <0.01&$+$0.01&0.02 & $\pm$0.01 & <0.01 & <0.01 & <0.01&$+$0.01&0.01\\
  \ion{K}{i} & $^{+0.06}_{-0.05}$ & $^{<0.01}_{+0.01}$ & $^{<0.01}_{+0.01}$ & $^{-0.01}_{+0.02}$&$+$0.04&0.06 & $\pm$0.06 & $^{<0.01}_{+0.01}$ & $^{<0.01}_{+0.01}$ & $^{-0.02}_{+0.03}$&$+$0.04&0.07\\
  \ion{Ca}{i} & $^{+0.04}_{-0.06}$ & $^{-0.02}_{+0.01}$ & $^{-0.01}_{<0.01}$ & $^{-0.02}_{+0.01}$&$+$0.03&0.06 & $^{+0.04}_{-0.05}$ & $^{-0.01}_{+0.02}$ & $\mp$0.01 & $\mp$0.02&$+$0.05&0.05\\
  \ion{Sc}{ii} & $^{+0.01}_{-0.02}$ & $\pm$0.03 & $\pm$0.01 & $\mp$0.02&$-$0.03&0.04 & $^{<0.01}_{-0.01}$ & $^{+0.02}_{-0.03}$ & $\pm$0.01 & $^{-0.02}_{+0.01}$&$-$0.05&0.03\\
  \ion{Ti}{i} & $^{+0.07}_{-0.08}$ & $\mp$0.02 & $^{-0.02}_{+0.01}$ & $\mp$0.02&$+$0.05&0.08 & $^{+0.09}_{-0.10}$ & $\mp$0.01 & $^{-0.02}_{+0.01}$ & $^{-0.03}_{+0.02}$&$+$0.03&0.10\\
  \ion{Ti}{ii} & $\pm$0.01 & $\pm$0.03 & $^{+0.01}_{<0.01}$ & $^{-0.01}_{+0.02}$&$-$0.03&0.04 & <0.01 & $\pm$0.03 & $\pm$0.01 & $\mp$0.01&$-$0.05&0.04\\
  \ion{V}{i} & $\pm$0.07 & $^{-0.02}_{+0.03}$ & $\mp$0.01 & $^{<0.01}_{+0.01}$&$+$0.05&0.08 & $\pm$0.09 & $^{-0.02}_{+0.03}$ & $\mp$0.02 & $^{<0.01}_{+0.01}$&$+$0.07&0.10\\
  \ion{Cr}{i} & $\pm$0.07 & $^{-0.01}_{+0.02}$ & $\mp$0.01 & $\mp$0.01&$+$0.04&0.07 & $\pm$0.08 & $\mp$0.02 & $^{-0.01}_{+0.02}$ & $\mp$0.03&$+$0.06&0.09\\
  \ion{Mn}{i} & $^{+0.06}_{-0.05}$ & $^{-0.01}_{+0.02}$ & $\mp$0.01 & $\mp$0.01&$+$0.04&0.06 & $^{+0.05}_{-0.06}$ & $^{-0.02}_{+0.03}$ & $^{-0.02}_{+0.01}$ & $\mp$0.01&$+$0.07&0.07\\
  \ion{Fe}{i} & $^{+0.07}_{-0.06}$ & $\mp$0.01 & $\mp$0.01 & $^{-0.02}_{+0.03}$&$+$0.04&0.07 & $\pm$0.06 & $\mp$0.01 & $\mp$0.01 & $\mp$0.03&$+$0.05&0.07\\
  \ion{Fe}{ii} & <0.01 & $\pm$0.03 & $\pm$0.01 & $^{-0.02}_{+0.03}$&$-$0.02&0.04 & $^{-0.02}_{+0.01}$ & $\pm$0.03 & $\pm$0.01 & $\mp$0.02&$-$0.05&0.04\\
  \ion{Co}{i} & $\pm$0.08 & $^{-0.03}_{+0.04}$ & $^{-0.01}_{+0.02}$ & $^{-0.05}_{+0.06}$&$+$0.09&0.11 & $\pm$0.08 & $\mp$0.03 & $\mp$0.02 & $^{-0.06}_{+0.07}$&$+$0.09&0.11\\
  \ion{Ni}{i} & $^{+0.07}_{-0.06}$ & $^{<0.01}_{+0.01}$ & $^{<0.01}_{+0.01}$ & $^{<0.01}_{+0.01}$&$+$0.04&0.07 & $^{+0.06}_{-0.07}$ & $\mp$0.01 & $^{-0.01}_{<0.01}$ & $\mp$0.01&$+$0.03&0.07\\
  \ion{Zn}{i} & $\pm$0.01 & $^{+0.02}_{-0.01}$ & <0.01 & $\mp$0.01&$-$0.01&0.02 & $\mp$0.01 & $\pm$0.01 & <0.01 & $\mp$0.02&$-$0.01&0.02\\
  \ion{Y}{ii} & $\pm$0.02 & $^{+0.03}_{-0.02}$ & $^{+0.01}_{<0.01}$ & $\mp$0.01&$-$0.03&0.03 & $^{+0.01}_{<0.01}$ & $\pm$0.02 & $^{+0.01}_{<0.01}$ & $^{-0.02}_{+0.03}$&$-$0.03&0.03\\
  \ion{Zr}{ii} & $\pm$0.02 & $\pm$0.02 & $\pm$0.01 & $\mp$0.03&$<$0.01&0.04 & $^{<0.01}_{-0.01}$ & $^{-0.01}_{<0.01}$ & <0.01 & $\mp$0.04&$<$0.01&0.04\\  
  \ion{Ba}{ii} & $\pm$0.03 & $\pm$0.03 & $^{+0.02}_{-0.01}$ & $^{-0.06}_{+0.07}$&$-$0.03&0.08 & $^{+0.02}_{-0.03}$ & $^{+0.04}_{-0.05}$ & $^{<0.01}_{-0.01}$ & $^{-0.08}_{+0.07}$&$-$0.08&0.09\\
  \ion{La}{ii} & $^{+0.03}_{-0.02}$ & $\pm$0.02 & $^{+0.01}_{<0.01}$ & $^{-0.02}_{+0.03}$&$-$0.02&0.04 & $\pm$0.01 & $^{+0.01}_{<0.01}$ & <0.01 & $\mp$0.01&$-$0.01&0.02\\
  \ion{Nd}{ii} & $\pm$0.03 & $^{+0.02}_{-0.03}$ & $^{<0.01}_{-0.01}$ & $\mp$0.01&$-$0.03&0.04 & $\pm$0.02 & $\pm$0.02 & $\pm$0.01 & $\mp$0.01&$-$0.04&0.03\\
  \ion{Sm}{ii} & ... & ... & ... & ...&...&... & $\pm$0.02 & $\pm$0.02 & $\pm$0.01 & <0.01&$-$0.04&0.03\\
  \ion{Eu}{ii} & $^{+0.03}_{-0.02}$ & $\pm$0.02 & $^{+0.01}_{<0.01}$ & $^{<0.01}_{+0.01}$&$-$0.02&0.03 & $\pm$0.01 & $^{+0.01}_{<0.01}$ & <0.01 & $^{<0.01}_{+0.01}$&$<$0.01&0.01\\
  \hline
  \end{tabular}}
  \end{table*}
  
  Our final statistical uncertainty $\sigma_\mathrm{stat}$ for the abundance of a species was computed from the $1\sigma$ rms line-to-line scatter. Consequently, this uncertainty accounts for spectral noise as well as errors of the atomic data. Once again, we clipped those lines that deviate more than $2.5\sigma_\mathrm{stat}$ from the weighted mean of all lines belonging to that species. For those cases, where only one line was detectable, we provide the error based on the $EW$ uncertainty, which has been propagated through the analysis. Finally, the total error budget $\sigma_\mathrm{tot}$ reported in Table \ref{Table:AB_RESI} represents the quadratic sum of $\sigma_\mathrm{stat}$ and $\sigma_\mathrm{atm}$.

  \subsection{Notes on individual elements}\label{Sec:indiv_elements}
\begin{figure}
  \centering
  \resizebox{\hsize}{!}{\includegraphics{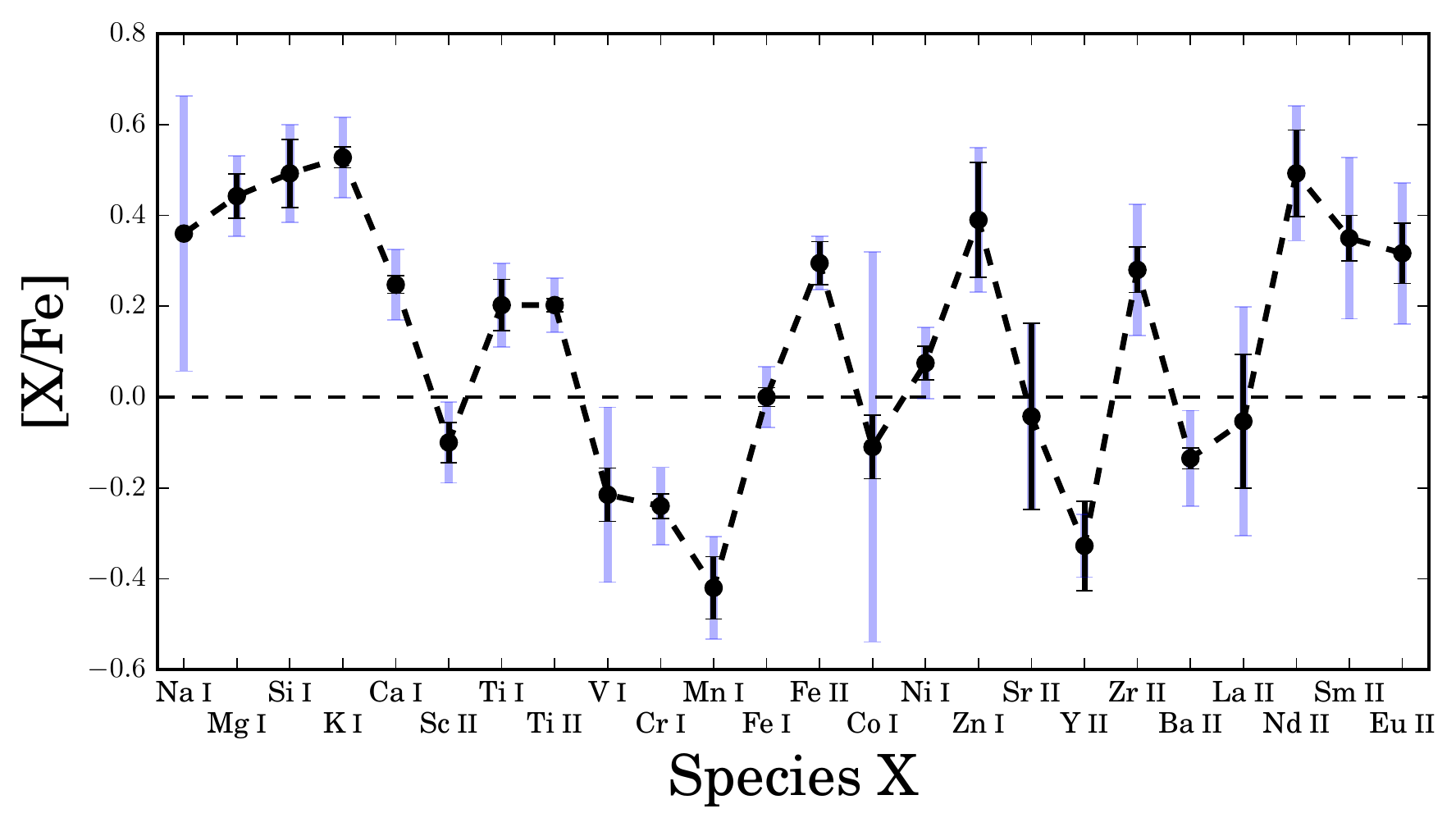}}
    \caption{Mean abundance results of all elements analyzed in this study. All ratios of neutral and ionized species are scaled to their Fe counterpart. The only exceptions made are \ion{Fe}{i} and \ion{Fe}{ii}, which are given with respect to the cluster mean and \ion{Fe}{i} abundances, respectively. The black error bars indicate the rms scatter around the mean, while the light blue error bars represent the mean $\sigma_\mathrm{tot}$ including stellar parameter uncertainties.    
      }
    \label{Fig:ALL_SPECIES_SOLAR}
  \end{figure}

  In the following sections, we will comment on individual results of our abundance measurements. Fig. \ref{Fig:ALL_SPECIES_SOLAR} shows a compilation of the mean cluster abundance ratios of all elements analyzed here.     
  
    \subsubsection{Fe}\label{Sec:iron}
\begin{figure}
    \centering
    \resizebox{\hsize}{!}{\includegraphics{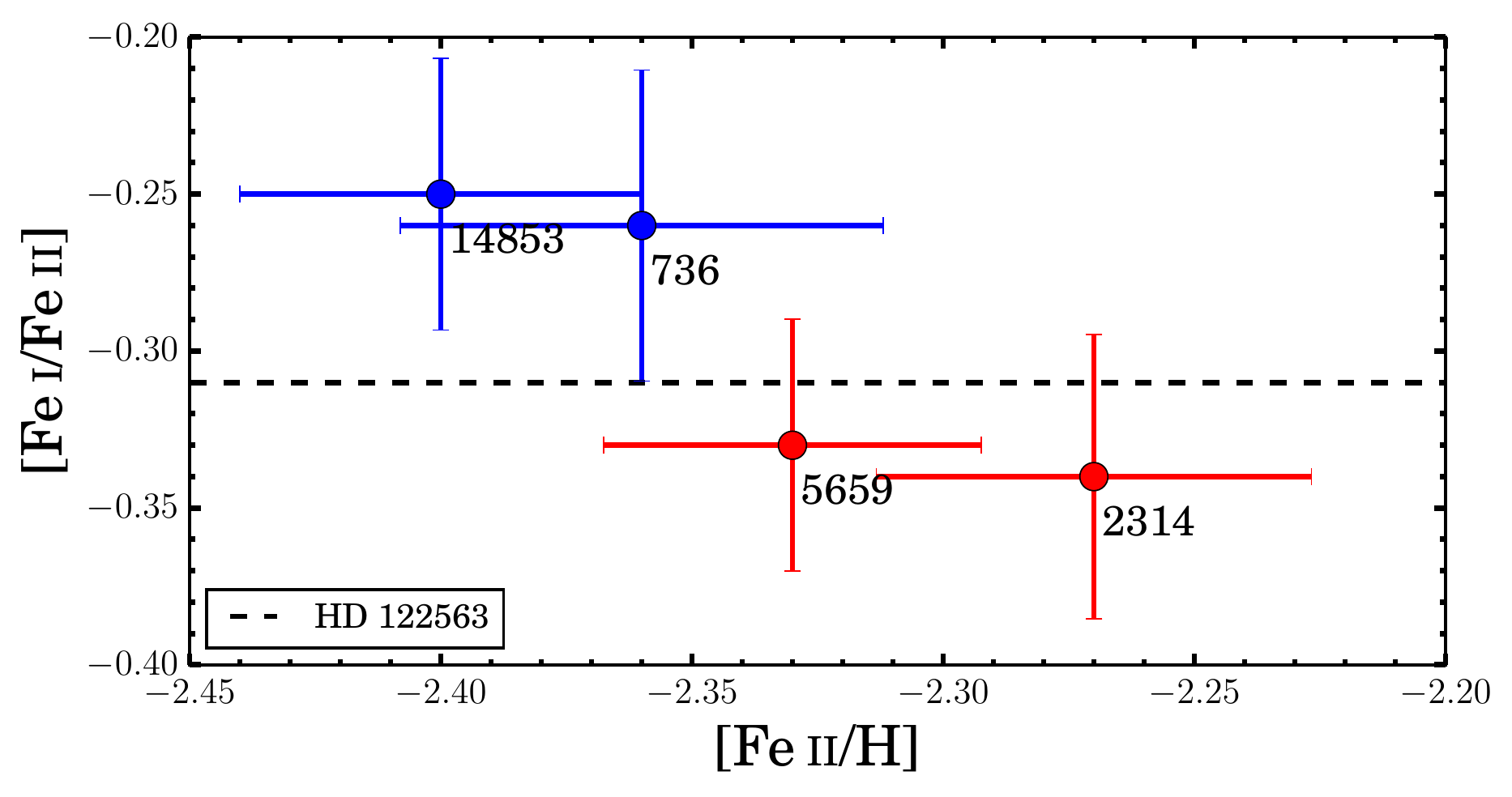}}
      \caption{Ionization imbalance between \ion{Fe}{i} and \ion{Fe}{ii}. Color coding of the stars was chosen to represent the membership of the potential sub-populations MP (blue) and MR (red). The indicated errors were computed by dividing the line-to-line abundance scatter by $\sqrt{N_\mathrm{lines}}$. The dashed line represents the [\ion{Fe}{i}/\ion{Fe}{ii}] found for HD~122563.
              }
      \label{Fig:Iron}
    \end{figure}

    Our final mean [Fe/H] for NGC~6426 is $-2.34$ dex with a scatter of 0.05 dex as derived from \ion{Fe}{ii} lines. When looking at the ratios of abundances of neutral to ionized iron species, [\ion{Fe}{i}/\ion{Fe}{ii}], we found a spurious anti-correlation with [\ion{Fe}{ii}/H], indicated in Fig. \ref{Fig:Iron}. 
    There, we highlight two groups of two stars each by color. One group appears to be deficient in [\ion{Fe}{ii}/H] by 0.08 dex and exposes an ionization imbalance [\ion{Fe}{i}/\ion{Fe}{ii}] elevated by $\sim+0.08$ dex compared to the other group, which tends to higher \ion{Fe}{ii} abundances and lower ionization imbalances. These imbalance deviations cannot be assigned to NLTE-effects, because either group contains both a mildly warmer, higher surface gravity star (14853/5659) and a cooler, lower surface gravity star (736/2314). Thus departures from LTE could be held responsible for an intra-group scatter of [\ion{Fe}{i}/\ion{Fe}{ii}] but not for the presented inter-group deviation. We can only speculate about the origin of this imbalance. If it is real, a possible explanation could be a significantly deviating He composition in both groups, which ultimately affects the model surface gravity and therefore the \ion{Fe}{ii} abundance \citep[see][for a detailed discussion]{Stromgren82,Lind11}. On the other hand, we encountered rather large uncertainties in the photometry and extinction maps, which could have a systematic effect on the surface gravities and as a consequence on the \ion{Fe}{ii} abundances, so that the small separation could also be linked to these. Owing to these doubts, at this point we can neither accept nor reject the hypothesis that we have actually observed two distinct populations in NGC~6426. We will discuss these and further tentative evidence from other elements in Sect. \ref{Sec:mult_pops}.
    
    \subsubsection{Alpha elements Mg, Si, Ca, Ti}
\begin{figure*}
    \centering
    \includegraphics[width=17cm]{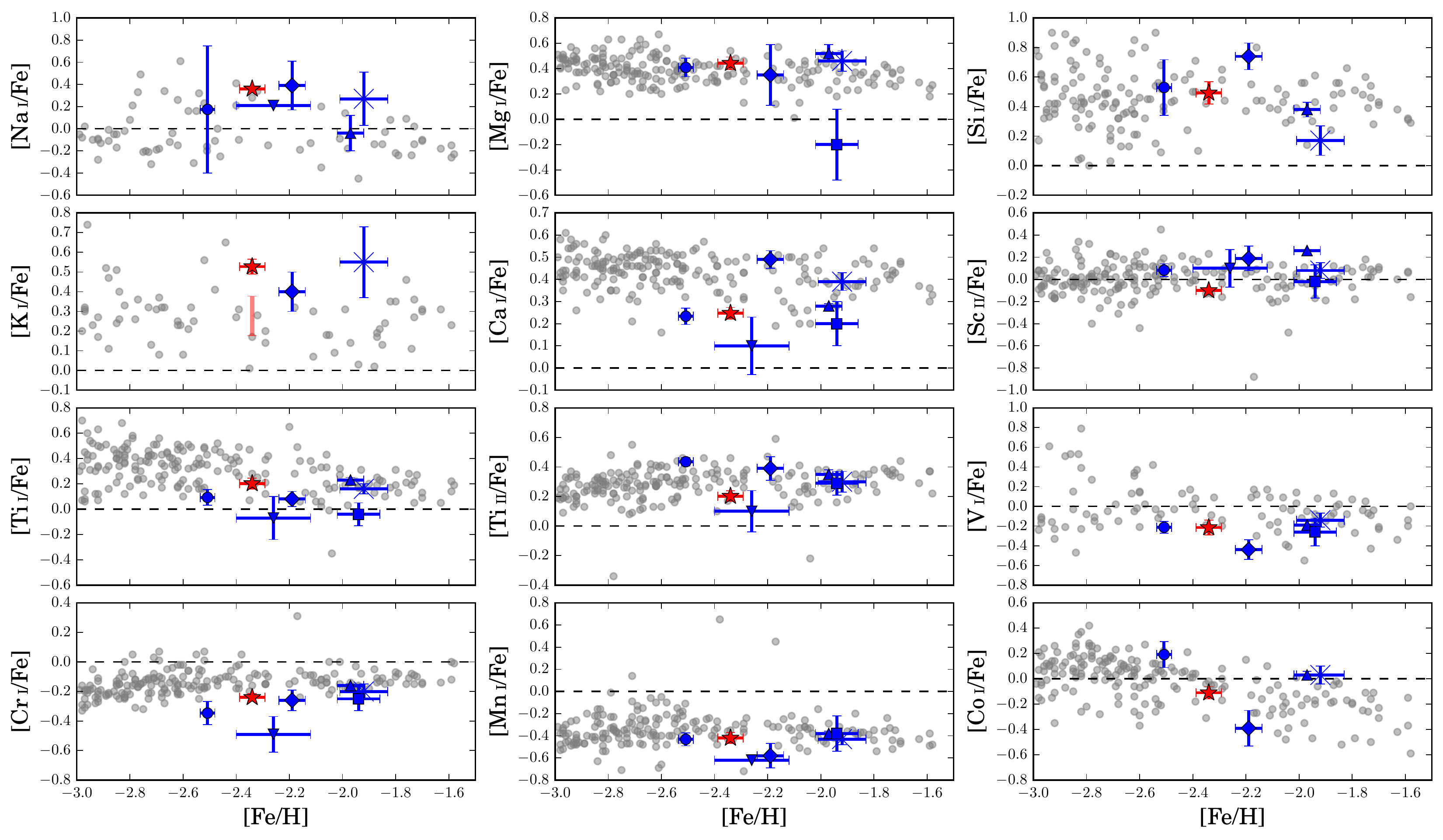}
      \caption{Mean results and scatter for [X/Fe] for all species detected with $11\leq Z \leq27$. NGC~6426 (this study) is represented by red stars. For the case of K, the red bar indicates the region where [K/Fe] is to be expected after an NLTE correction. The gray points resemble the compilation of abundances of stars in the MW halo by \citet{Roederer14} in the range $-3$ dex $\leq$ [Fe/H] $\leq -1.5$ dex. We also include (in blue) the mean abundances of the similarly metal-poor GCs NGC~5897 \citep[crosses,][]{Koch14}, NGC~4833 \citep[diamonds,][]{Roederer15}, NGC~5824 \citep[squares,][]{Roederer16}, M15 \citep[filled circles, RGB targets of][]{Sobeck11}, NGC~5634 and NGC~5053 \citep[up and down triangles,][]{Sbordone15}.  
              }
      \label{Fig:HALO_AND_OTHERS_I}
    \end{figure*}

    In terms of light $\alpha$-elements, we found [Mg/Fe$]=0.44\pm0.05$ dex, confirming previous results from low-resolution spectroscopy \citep[0.38$\pm$0.06 dex,][]{Dias15}, and --for the first time-- [Si/Fe$]=0.49\pm0.07$ dex. These two species will be discussed in more detail in Sect. \ref{Sec:mult_pops}. For the heavy $\alpha$-elements Ca and Ti in NGC~6426 we report mean abundances and rms scatters of $0.25\pm0.03$ dex and $0.20\pm0.04$ dex for [Ca/Fe] and [Ti/Fe], respectively. Hence, the cluster falls in line with metal-poor halo stars at the same metallicity (see Fig. \ref{Fig:HALO_AND_OTHERS_I}), which show elevated levels of pure $\alpha$- and Ti abundances \citep[see][]{Pritzl05,Koch14}. There is no intrinsic separation in these two elements and the scatter is fully consistent with the uncertainties. Our analysis reveals a mean [Mg/Ca] that is enhanced by 0.19~dex. Yield models for massive core-collapse supernovae (SNe) -- with progenitor masses of $15M_\sun$ and above-- predict enhancements of [Mg/Ca] \citep[see, e.g.,][and references therein]{Koch09a}. Consequently, a possible explanation for our finding is the imprint of nucleosynthesis by such SN events (see Sec. \ref{Sec:heavy_el} and Fig. \ref{Fig:SN_vs_HN} for further constraints and discussion).
    
    \begin{figure*}
    \centering
    \includegraphics[width=17cm]{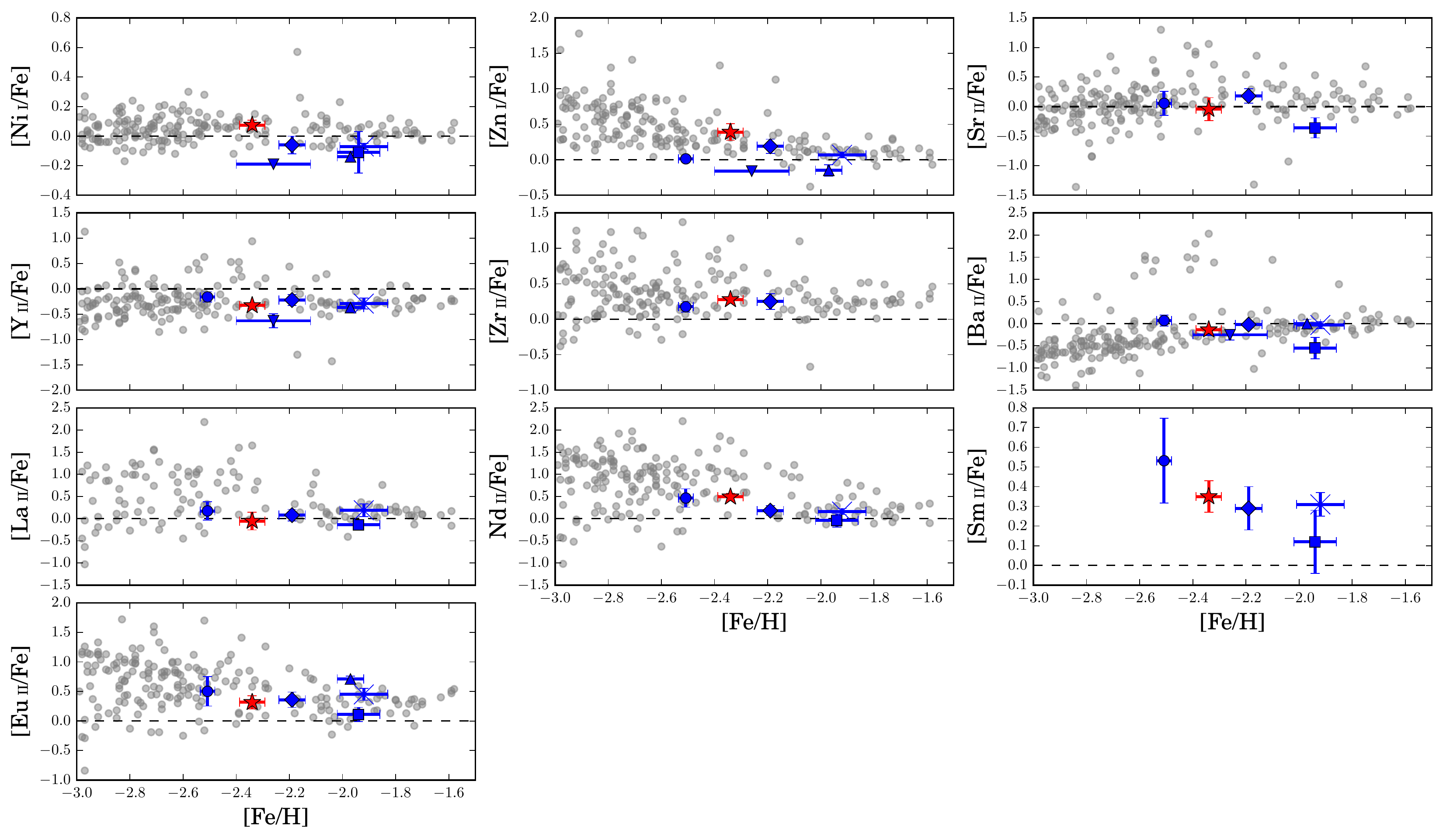}
      \caption{Same as in Fig. \ref{Fig:HALO_AND_OTHERS_I} for $28\leq Z \leq63$.
              }
      \label{Fig:HALO_AND_OTHERS_II}
    \end{figure*}
    
    \subsubsection{Light elements Na, O, Al, K}
    Unfortunately, due to limited spectrum quality and the metal-poor nature of our sample we were not able to derive O and Al abundances using either EW measurements or with spectrum synthesis. For Na we were only able to measure one line for the star 5659. An NLTE-corrected abundance for this line has been computed from \citet{Lind11Na}\footnote{Data obtained from the INSPECT database, version 1.0 (www.inspect-stars.net)}. The NaD lines were too strongly saturated and affected by interstellar absoption for all targets. The existence or non-existence of the popular Na-O and Mg-Al anti-correlations \citep{Carretta09, Gratton12} could be used to put further constraints on the multiple population hypothesis.
    
    Our LTE analysis of the \ion{K}{i} resonance line at 7698.97 {\AA} shows [K/Fe$]\sim 0.53$ dex. However this has to be treated with caution, since for this particular line strong NLTE-corrections can be expected as stated by \citet{Takeda09}. From their work, we expect corrections between $-0.2$ and $-0.3$ dex in the regime of our derived stellar parameters. Consequently NGC~6426's K abundances fall on the trend of the GC M15 in \citet{Takeda09} and the MW halo at comparable metallicities (see again Fig. \ref{Fig:HALO_AND_OTHERS_I}).

    \subsubsection{Iron peak elements Sc, V, Cr, Mn, Co, Ni}
    Concerning the iron peak, given its metallicity we can report on a regular behavior of NGC~6426. Our study reveals little scatter among the sample stars for the iron peak element abundances. The rms scatter values range from 0.01 dex for [Sc/Fe] to 0.05 dex  for [Mn/Fe], which is well within the respective individual uncertainties. Again, we found the cluster mean abundances for Sc, V, Cr, Mn, Co and Ni to coincide well with other GCs and the halo at comparable [Fe/H] (see Figs. \ref{Fig:HALO_AND_OTHERS_I} and \ref{Fig:HALO_AND_OTHERS_II}). 
    
    \subsubsection{Neutron-capture elements Zn, Sr, Y, Zr, Ba, La, Nd, Sm, Eu}\label{Sec:heavy_el}
    The mean [Zn/Fe] ratio for NGC~6426 appears to be enhanced by 0.39 dex and falls on the observed trend of increasing [Zn/Fe] with decreasing [Fe/H] of galactic halo field stars first reported by \citet{Johnson99}. Possible production sites of Zn are the complete Si-burning layers of massive stars \citep{Umeda02}. In these layers, Si is completely destroyed. \citet{Umeda05} concluded that observed supersolar [Zn/Fe] in metal-poor halo stars imply an enrichment by very energetic massive supernovae (the so-called hypernovae, HN). In their models of such events, the fraction of complete Si-burning layer (high Zn production rate) to incomplete Si-burning layer (low Zn production rate) is elevated, such that the abundance observations are matched. 
    \begin{figure*}
    \centering
    \includegraphics[width=17cm]{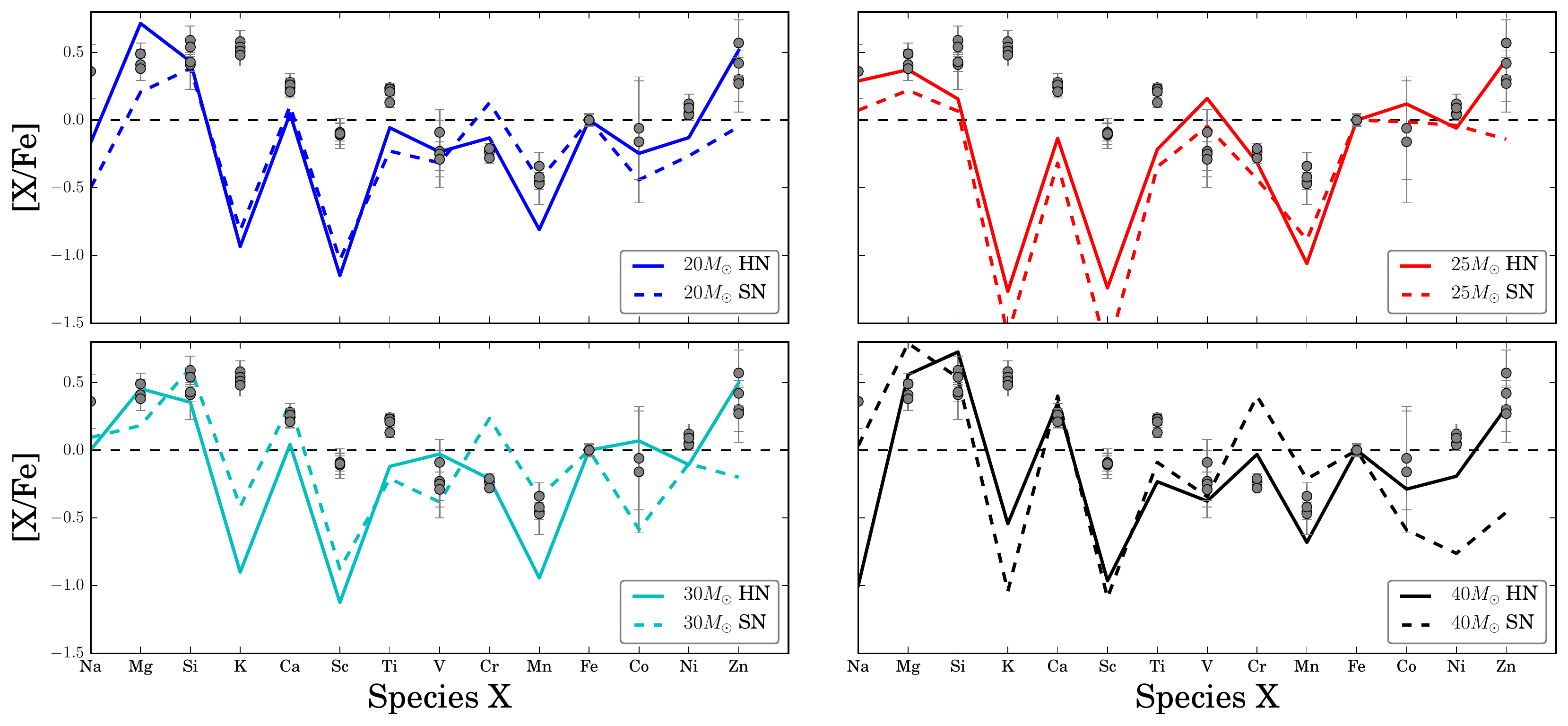}
      \caption{Model yields from \citet{Nomoto13} for $Z=0.001$, different SN explosion energies, and progenitor masses compared to this study's individual stellar abundances for the elements Na to Zn.
              }
      \label{Fig:SN_vs_HN}
    \end{figure*}
    Figure \ref{Fig:SN_vs_HN} shows a comparison of predicted yields for normal supernovae and HNe models with assumed explosion energies ten times larger than in normal core-collapse SNe. The respective yields for progenitor masses of $20M_\sun,$ $25M_\sun,$ $30M_\sun$ and $40M_\sun$ were taken from \citet{Nomoto13} and are presented along with our findings for the elements from Na to Zn. Apparently, normal SN fail to reproduce the overall trend from Fe to Zn, especially the Ni-Zn relation. In terms of minimum $\chi^2$, the best global fit is achieved by the $40M_\sun$ HN model, whereas a $25M_\sun$ HN seems to resemble the Fe to Zn pattern best because of the predicted close to solar [Ni/Fe]. Hence it is feasible that the pristine gas, from which NGC~6426 evolved, has experienced HN enrichment. 
    
    As shown in previous studies comparing observations to a number of different yield predictions \citep[e.g.,][]{Hansen11} finding a perfect match to SN or HN yields can be very challenging even for the less polluted extremely metal-poor stars. A number of factors cloud such comparisons, such as the number of free parameters in the model predictions (mass, energy, number of electrons per nucleon, $T_\mathrm{peak}$, jets, fallback and the contribution from neutrino-driven winds). Also uncertainties in stellar abundances (systematics, abundance assumptions such as LTE and 1D) combined with various unknowns in ISM mixing add to the uncertainty in the yield comparison. Unless a perfect match to a given model is found, one model cannot be singled out as the only donor, but mixing of various contributions is likely to have taken place and changed the original composition of the yielded abundance ratios. At this metallicity ( $\sim-2.3$ dex) we are most likely not seeing the clean trace of one event and the somewhat poor match to the predictions confirms this. We therefore do not believe that strong constraints on the SN/HN origin can be obtained, however, quantitative predictions as to the mass (based on the high level of lighter alpha-elements and the enhancement of [Mg/Ca]) and the energy being high (high Ni and Zn abundances) can be made. The number of events can, to date, not be derived with confidence, owing to degeneracies like the number of free model parameters and the uncertainties in the elemental abundances.
    
    We were not able to reliably measure $EW$s of isolated lines of Sr, because of strong blending in the blue wavelength regime. Thus, we applied spectrum synthesis for the resonance feature of \ion{Sr}{ii} at 4215 {\AA} and adjusted the input abundance to achieve the best fit \citep[see][for a detailed discussion]{Hansen13}. Typical errors introduced this way are of order $\sim0.2$ dex.  
    
    Elements heavier than Zn (Z > 30, the elements mainly attributed to neutron-capture processes) do not appear to show any means of separation in sub-populations when considering uncertainties. A possible exception is La which exposes the largest star-to-star scatter among all elements measured here (0.2 dex). On the other hand, being based on only one highly uncertain $EW$, La abundances have a typical uncertainty of $\sim 0.25$ dex. We note that the abundances of heavy elements from La to Eu are consistent with an r-dominated enrichment history with little to no contribution of the main slow neutron-capture process (s-process) attributed to low-mass AGB stars. This finding is supported by a low [Ba/Eu] of -0.46 dex, setting the upper limit of the formation timescale of NGC~6426 to a few hundred Myr.
    
    The distribution of the light neutron-capture elements Sr, Y and Zr cannot be assigned to either of the r- and s-process patterns, because of the deep depression of Y between enhanced Sr and Zr. The same sequence has previously been reported for other GCs, for example, for M15 by \citet{Sobeck11} and partly for NGC~5897 (low Y) by \citet{Koch14}. Both groups assign these findings to the contribution of a yet unknown lighter element primary process \citep[LEPP,][]{McWilliam98,Travaglio04}, which ought to be active in metal-poor environments. Low Y abundances have also been reported by \citet{Hansen12} for a large sample of halo stars.
    
  \subsection{Indications for multiple populations}\label{Sec:mult_pops}
  During our analysis we found several weak indications for a separation in two subpopulations within the sample of four NGC~6426 stars. None of the following individual arguments is significant on a $3\sigma$-level -- but taken together, they draw a consistent, yet still uncertain picture.
  
  The first clue is the already discussed ionization imbalance among the Fe ionization states (Sect. \ref{Sec:iron}). Two-tailed t-tests of the separations in the two dimensions [\ion{Fe}{ii}/H] and [\ion{Fe}{i}/\ion{Fe}{ii}] revealed statistical significances of 89.2\% and 99.6\%, respectively. Hence, at this point, the null hypothesis of actually having observed only one population cannot be rejected in neither of these abundance ratios. Nevertheless, for labeling purposes, we will from now on refer to the (mildly) more metal-poor (MP, stars 14853 and 736) and (mildly) more metal-rich population (MR, stars 5659 and 2314), when elaborating on their distinct nature.
  
\begin{figure}
      \centering
      \resizebox{\hsize}{!}{\includegraphics{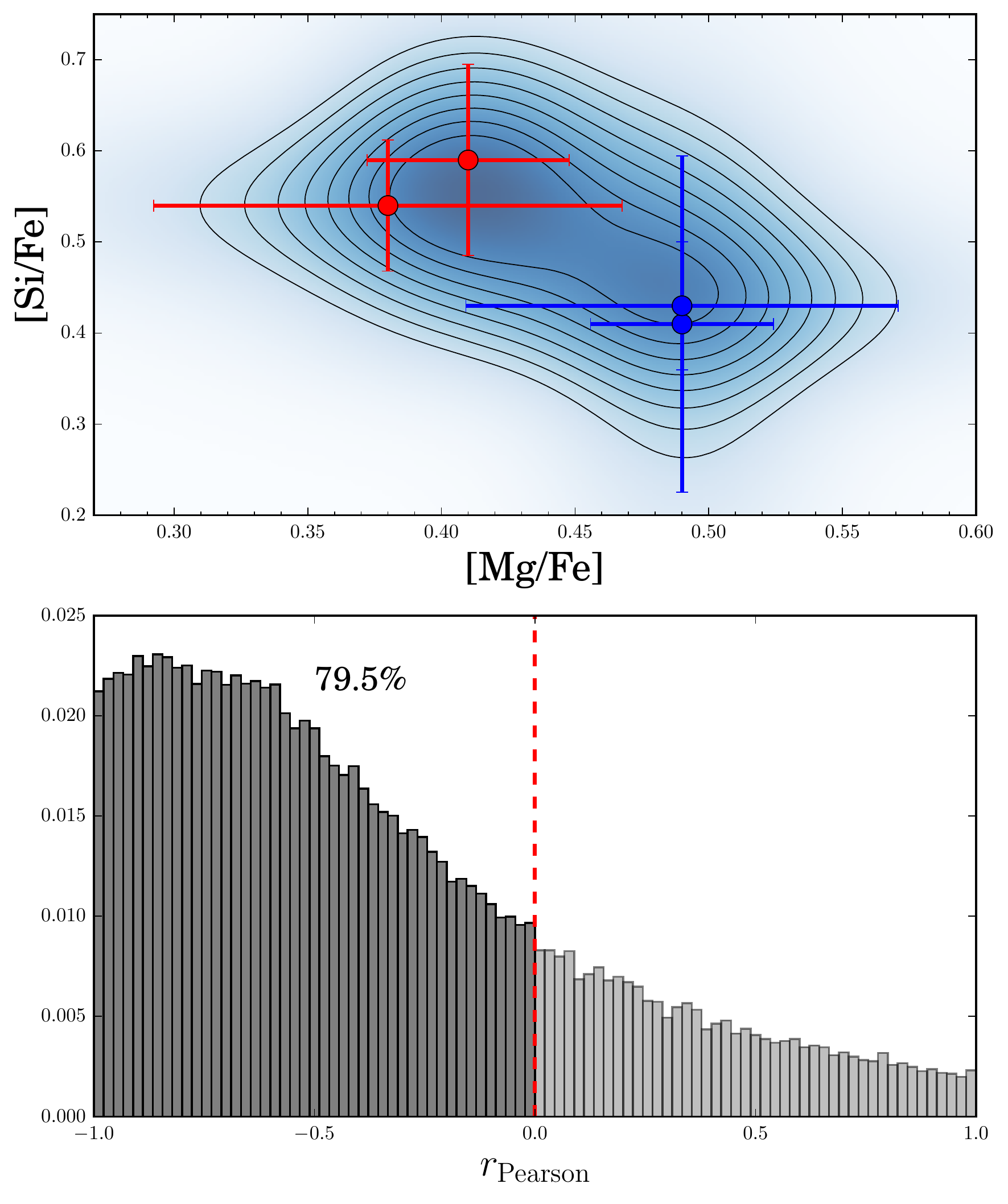}}
      \caption{Top: Abundance results for Si and Mg. Color coding is the same as in Fig. \ref{Fig:Iron}. The contours show the smoothed distribution, which was used in order to evaluate the significance of the correlation. Bottom: Discrete probability distribution of the correlation coefficients for the $10^5$ realizations. 
              }
      \label{Fig:MG_SI}
  \end{figure}

  The abundance results for Mg are consistent within the given uncertainties, though we note that both stars, which we previously attributed to the MP population expose the more enhanced [Mg/Fe] compared to the MR population. There is evidence of a more pronounced separation when looking at the [Si/Fe] ratios. The MP stars expose an enhancement of 0.41 and 0.43 dex, while the MR stars seem to be even more enhanced by additional 0.15 dex (absolute 0.59 and 0.54 dex). There is a weak indication for a Mg-Si anti-correlation with a correlation coefficient of -0.89. In order to get a handle on the significance of this anti-correlation and to account for our large uncertainties, we randomly generated $10^5$ sets of four points in the [Mg/Fe]-[Si/Fe] plane where each point was drawn from a normal distribution resembling the means and uncertainties of the underlying cluster star. For each of the sets we computed the Pearson correlation coefficient $r_\mathrm{Pearson}$. The result is presented in Fig. \ref{Fig:MG_SI}.
  The lower panel shows a histogram from which we conclude that $79.5\%$ of all realizations actually are anti-correlations. Thus we excluded the cases correlation ($r>0$) and no correlation at all ($r=0$) with a confidence level of $79.5\%$. This is far from enabling us to claim significance, but it is the same stars that already separated in [\ion{Fe}{i}/\ion{Fe}{ii}]. 
    
  Mg is depleted in the H-burning phases of hot and massive stars, where the Mg-Al cycle is active. There, Mg gets destroyed at the expense of producing Al through a series of proton captures and successive $\beta$-decays \citep[see, e.g.,][]{Karakas03}. Since the circle is not entirely closed, Si can be produced by further proton captures on Al seed nuclei. Therefore a depletion in Mg as a result of this process should come with an enhancement of Si. \citet{Carretta09b} proposed this ``leakage'' in the Mg-Al chain to be responsible for their observations of Mg-Si anti-correlations in metal-poor and/or massive clusters. Consequently, assuming its existence, the observed anti-correlation in NGC~6426 points toward the two population scenario, where the ejecta of evolved, massive stars are depleted in Mg and enhanced in Si, which affects the composition of the ISM and eventually the newly formed stars.
  
\begin{figure}
    \centering
    \resizebox{\hsize}{!}{\includegraphics{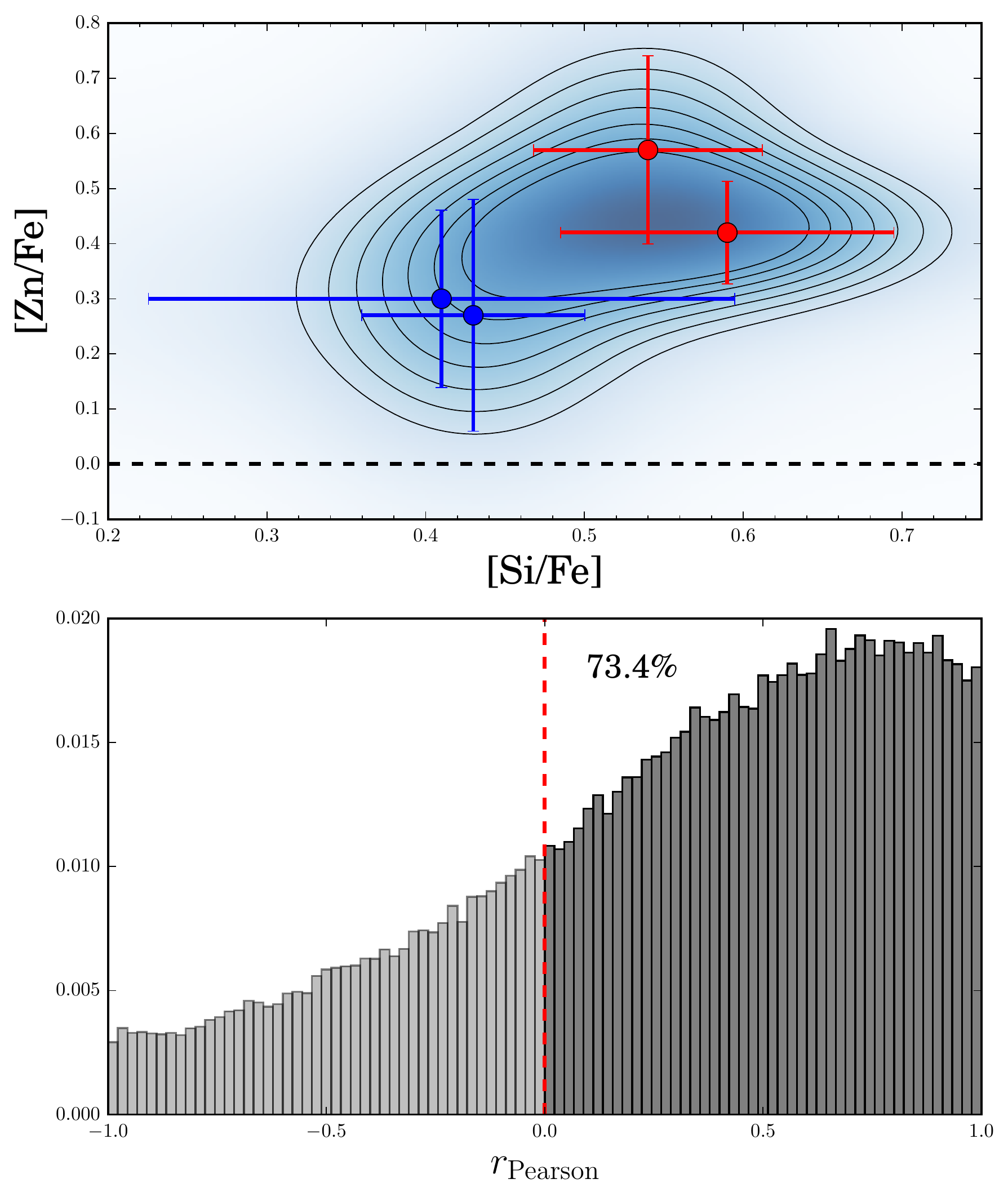}}
      \caption{Same as in Fig. \ref{Fig:MG_SI} but for Zn and Si.
              }
      \label{Fig:SI_ZN}
    \end{figure}

  The rms scatter in [Zn/Fe] of our four stars is lower than the individual uncertainties. Nevertheless, the left panel of Fig. \ref{Fig:SI_ZN} once again indicates a systematic separation of $\sim 0.2$ dex between MP and MR stars. Moreover, we report on a weak, non-significant correlation of [Zn/Fe] with [Si/Fe] ($r_\mathrm{Pearson}=0.75$, confidence level $73.4\%$, see Fig. \ref{Fig:SI_ZN}).
  If the intra-cluster correlation of [Zn/Fe] with [Si/Fe] is real, it further supports the argument that the enhancement of [Si/Fe] between MP and MR stars originates from pollution by hot and massive MP stars. In those stars with masses above $15M_\sun$, the weak component of the s-process is active. This process is believed to be a production site of elements between Fe and Sr such as Zn \citep{Kappeler89,Frischknecht12}. Consequently both, elevated [Zn/Fe] and higher [Si/Fe] levels of MR stars point toward an additional contribution from massive star nucleosynthesis --  weak s-process and hot H-burning cycles -- to the element budget of these stars. Taking the aforementioned scenarios for granted, the MP population should resemble the unpolluted HN yield patterns presented in Fig. \ref{Fig:SN_vs_HN} more closely than the MR population. In fact, the fit quality of the $40M_\sun$ HN model for the global pattern and the $25M_\sun$ HN case for the Fe to Zn trend improves slightly, once we only consider the MP population. Nevertheless, we emphasize once again that the uncertainties involved remain high and the conclusions drawn here are highly speculative.
    
  \subsection{Similarity to HD~108317}
\begin{figure}
  \centering
  \resizebox{\hsize}{!}{\includegraphics{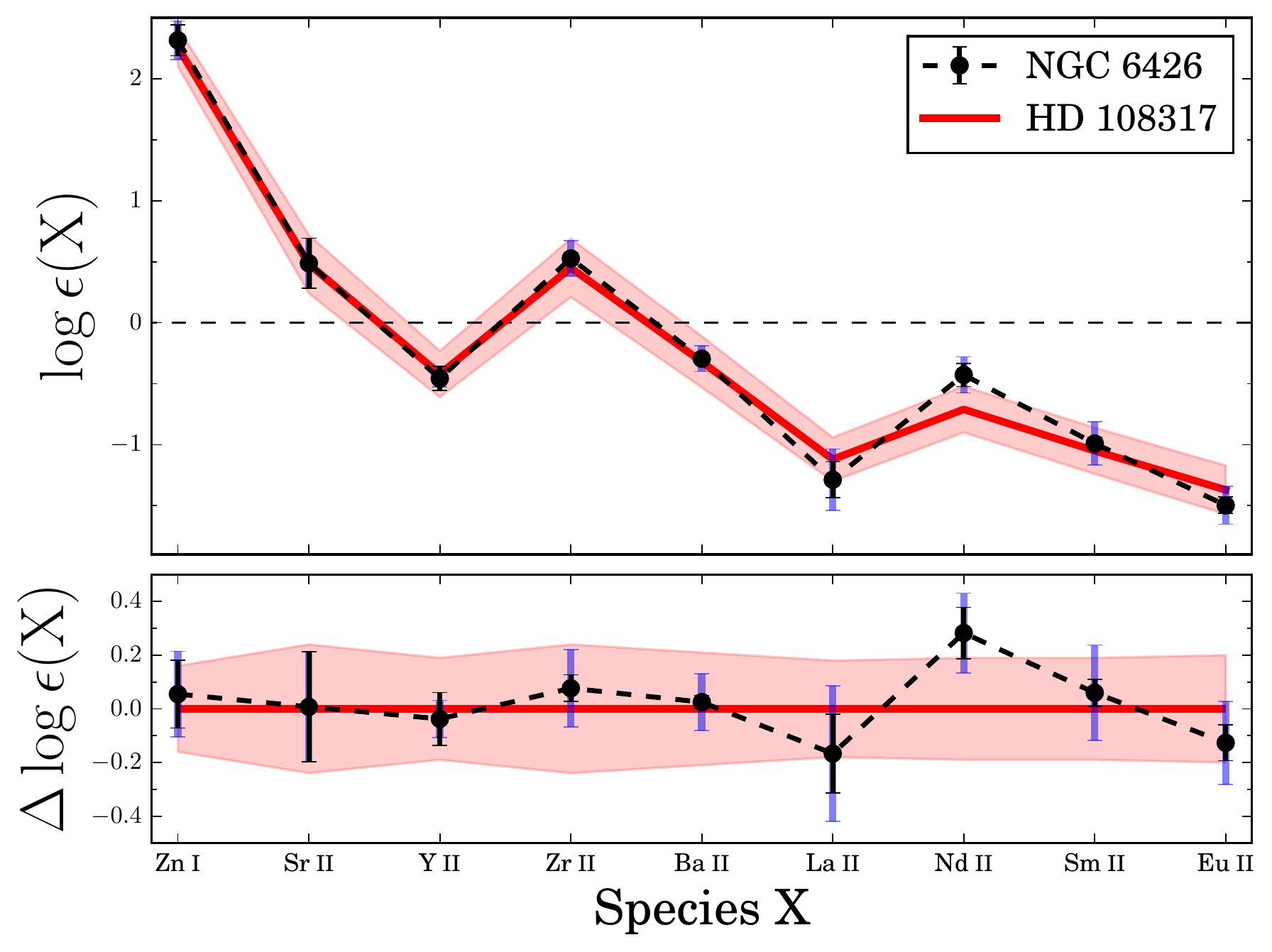}}
    \caption{Top panel: Mean abundance pattern for NGC~6426. The color coding is the same as in Fig. \ref{Fig:ALL_SPECIES_SOLAR}. The red line and red shaded region state the abundances and uncertainties of HD~108317 as derived by \citet{Roederer12}. Both patterns are unscaled, i.e., represent their original values. Bottom panel: Residual pattern for the cluster stars after subtracting HD~108317. 
              }
    \label{Fig:HEAVY_ELEMENTS}
  \end{figure}

  In terms of abundances of Zn and the neutron-capture elements, we note a remarkable overlap of our target stars with the analysis of HD~108317 by \citet{Roederer12}. Their study of this close-by halo red giant ($T_\mathrm{eff}=5100$ K, $\log{g}=2.67$ dex, $v_\mathrm{mic}=1.50$ km s$^{-1}$) revealed an Fe abundance of [\ion{Fe}{ii}/H$]=-2.37\pm0.14$ (stat.) dex, which falls right on top of our findings for NGC~6426. Its heavy element pattern in conjunction with the one of NGC~6426 is shown in Fig. \ref{Fig:HEAVY_ELEMENTS}. Neither of them have been scaled to fit the other, meaning both patterns are the original results. The Nd uncertainty does not change the results, since the combined uncertainties from both studies remain high. It is tempting to say that this striking congruence points to the conclusion that whatever processes contributed to the chemical enrichment of heavy elements in HD~108317, have also been active in NGC~6426. \citet{Roederer12} claim these to be mainly the r-process with a possible minority component of the s-process. In light of published abundance measurements of HD~108317 other than the aforementioned, however, the s-process might have had a more pronounced contribution. \citet{Hansen12} and \citet{Burris00}, for example, consistently report on [Ba/Eu$]=0.00$ dex and [Ba/Eu$]=-0.13$ dex, respectively, compared to [Ba/Eu$]=-0.61$ dex from \citet{Roederer12}. These results indicate a mixed r-/s- origin of this star which is supported by \citet{Simmerer04} who --based on their [La/Eu] ratio-- excluded pure r-process enrichment.  

\section{Summary}\label{Sec:summ}
We have carried out the first spectroscopic high-resolution study of four stars of the GC NGC~6426 using the MIKE spectrograph. In doing so, we have spectroscopically derived the stellar parameters of the targets and computed chemical abundances of 22 elements, including $\alpha$-, iron peak- and several neutron-capture elements. We derived a metallicity of [Fe/H$]=-2.34\pm0.05$ dex, which is in good agreement with the spectroscopic results of \citet[$-2.39\pm0.04$ dex]{Dias15} from low-resolution Ca triplet spectroscopy and the photometric ones by \citet[$-2.33\pm0.15$ dex]{Hatzidimitriou99}.

The pattern of elemental abundances of the cluster fits well into the picture drawn by the MW halo and other GCs at comparable metallicities. In particular, we found a mean $\alpha$ enhancement of [(Mg,Si,Ca)/3 Fe$]=0.39\pm0.03$ dex and heavy element distributions that both fall in line with the abundance trends of the MW halo. There are weak indications for correlations and anti-correlations among the elements Mg, Si and Zn. Assuming these trends are real, we suggest that they point toward a pollution by heavy star nucleosynthesis of an older population toward a slightly younger one. We report on an overall enhancement of [Zn/Fe] by 0.39 dex, possibly indicating an early enrichment of the pre-cluster medium by hypernova events. Our results for the heavier neutron-capture elements La, Nd, Sm and Eu point toward an enrichment history governed by the r-process. However, the light neutron-capture elements Sr, Y and Zr neither fit the typical r- or s-process patterns nor a combination of both, because of the depletion of [Y/Fe] and the enhancement of [Sr/Fe] and [Zr/Fe]. We follow the argumentation of previous authors and attribute the results to an additional, yet unknown, source of these elements, called LEPP.

In terms of heavy elements, we encountered an extraordinary similarity of the MW field star HD~108317 and our program stars. Despite the disagreement in literature about the amount of s-process contributions to the enrichment of HD~108317, we suggest that this star was born in an environment similar to NGC~6426's natal cloud.       

\begin{acknowledgements}
AK acknowledges the Deutsche Forschungsgemeinschaft for funding from Emmy-Noether grant  Ko 4161/1. CJH was supported by a research grant (VKR023371) from VILLUM FONDEN. We are grateful to the referee for helpful and constructive inputs. The authors thank Ian Roederer for providing the MIKE spectrum of HD~122563. This research made use of atomic data from the INSPECT database, version 1.0 (www.inspect-stars.net), the NASA's Astrophysics Data System and the SIMBAD database, operated at CDS, Strasbourg, France. Moreover, we applied results from the NASA/ IPAC Infrared Science Archive, which is operated by the Jet Propulsion Laboratory, California Institute of Technology, under contract with the National Aeronautics and Space Administration. Atomic data were partly retrieved from the VALD database, operated at Uppsala University, the Institute of Astronomy RAS in Moscow, and the University of Vienna. 
\end{acknowledgements}

\bibliographystyle{aa.bst}
\bibliography{sources.bib}

\end{document}